\documentclass[useAMS,usenatbib]{mn2e}
\include{epsf}
\newcommand\aj{{AJ}}%
\newcommand\araa{{ARA\&A}}%
\newcommand\apj{{ApJ}}%
\newcommand\apjl{{ApJ}}%
\newcommand\apjs{{ApJS}}%
\newcommand\ao{{Appl.~Opt.}}%
\newcommand\apss{{Ap\&SS}}%
\newcommand\aap{{A\&A}}%
\newcommand\aapr{{A\&A~Rev.}}%
\newcommand\aaps{{A\&AS}}%
\newcommand\azh{{AZh}}%
\newcommand\baas{{BAAS}}%
\newcommand\jrasc{{JRASC}}%
\newcommand\memras{{MmRAS}}%
\newcommand\mnras{{MNRAS}}%
\newcommand\pra{{Phys.~Rev.~A}}%
\newcommand\prb{{Phys.~Rev.~B}}%
\newcommand\prc{{Phys.~Rev.~C}}%
\newcommand\prd{{Phys.~Rev.~D}}%
\newcommand\pre{{Phys.~Rev.~E}}%
\newcommand\prl{{Phys.~Rev.~Lett.}}%
\newcommand\pasp{{PASP}}%
\newcommand\pasj{{PASJ}}%
\newcommand\qjras{{QJRAS}}%
\newcommand\skytel{{S\&T}}%
\newcommand\solphys{{Sol.~Phys.}}%
\newcommand\sovast{{Soviet~Ast.}}%
\newcommand\ssr{{Space~Sci.~Rev.}}%
\newcommand\zap{{ZAp}}%
\newcommand\nat{{Nature}}%
\newcommand\iaucirc{{IAU~Circ.}}%
\newcommand\aplett{{Astrophys.~Lett.}}%
\newcommand\apspr{{Astrophys.~Space~Phys.~Res.}}%
\newcommand\bain{{Bull.~Astron.~Inst.~Netherlands}}%
\newcommand\fcp{{Fund.~Cosmic~Phys.}}%
\newcommand\gca{{Geochim.~Cosmochim.~Acta}}%
\newcommand\grl{{Geophys.~Res.~Lett.}}%
\newcommand\jcp{{J.~Chem.~Phys.}}%
\newcommand\jgr{{J.~Geophys.~Res.}}%
\newcommand\jqsrt{{J.~Quant.~Spec.~Radiat.~Transf.}}%
\newcommand\memsai{{Mem.~Soc.~Astron.~Italiana}}%
\newcommand\nphysa{{Nucl.~Phys.~A}}%
\newcommand\physrep{{Phys.~Rep.}}%
\newcommand\physscr{{Phys.~Scr}}%
\newcommand\planss{{Planet.~Space~Sci.}}%
\newcommand\procspie{{Proc.~SPIE}}%

\usepackage{amsmath}
\usepackage{graphicx}
\usepackage{subfig}
\usepackage{float}
\usepackage{upquote}
\usepackage{footnote}
\usepackage[T1]{fontenc}
\usepackage{aecompl}
\usepackage{auto-pst-pdf}

\def\jref@jnl#1{{\rm#1}}

\def\aj{\jref@jnl{AJ}}                   
\def\araa{\jref@jnl{ARA\&A}}             
\def\apj{\jref@jnl{ApJ}}                 
\def\apjl{\jref@jnl{ApJ}}                
\def\apjs{\jref@jnl{ApJS}}               
\def\ao{\jref@jnl{Appl.~Opt.}}           
\def\apss{\jref@jnl{Ap\&SS}}             
\def\aap{\jref@jnl{A\&A}}                
\def\aapr{\jref@jnl{A\&A~Rev.}}          
\def\aaps{\jref@jnl{A\&AS}}              
\def\azh{\jref@jnl{AZh}}                 
\def\baas{\jref@jnl{BAAS}}               
\def\jrasc{\jref@jnl{JRASC}}             
\def\memras{\jref@jnl{MmRAS}}            
\def\mnras{\jref@jnl{MNRAS}}             
\def\pra{\jref@jnl{Phys.~Rev.~A}}        
\def\prb{\jref@jnl{Phys.~Rev.~B}}        
\def\prc{\jref@jnl{Phys.~Rev.~C}}        
\def\prd{\jref@jnl{Phys.~Rev.~D}}        
\def\pre{\jref@jnl{Phys.~Rev.~E}}        
\def\prl{\jref@jnl{Phys.~Rev.~Lett.}}    
\def\pasp{\jref@jnl{PASP}}               
\def\pasj{\jref@jnl{PASJ}}               
\def\qjras{\jref@jnl{QJRAS}}             
\def\skytel{\jref@jnl{S\&T}}             
\def\solphys{\jref@jnl{Sol.~Phys.}}      
\def\sovast{\jref@jnl{Soviet~Ast.}}      
\def\ssr{\jref@jnl{Space~Sci.~Rev.}}     
\def\zap{\jref@jnl{ZAp}}                 
\def\nat{\jref@jnl{Nature}}              
\def\iaucirc{\jref@jnl{IAU~Circ.}}       
\def\aplett{\jref@jnl{Astrophys.~Lett.}} 
\def\apspr{\jref@jnl{Astrophys.~Space~Phys.~Res.}}
\def\bain{\jref@jnl{Bull.~Astron.~Inst.~Netherlands}} 
\def\fcp{\jref@jnl{Fund.~Cosmic~Phys.}}  
\def\gca{\jref@jnl{Geochim.~Cosmochim.~Acta}}   
\def\grl{\jref@jnl{Geophys.~Res.~Lett.}} 
\def\jcp{\jref@jnl{J.~Chem.~Phys.}}      
\def\jgr{\jref@jnl{J.~Geophys.~Res.}}    
\def\jqsrt{\jref@jnl{J.~Quant.~Spec.~Radiat.~Transf.}}
\def\memsai{\jref@jnl{Mem.~Soc.~Astron.~Italiana}}
\def\nphysa{\jref@jnl{Nucl.~Phys.~A}}   
\def\physrep{\jref@jnl{Phys.~Rep.}}   
\def\physscr{\jref@jnl{Phys.~Scr}}   
\def\planss{\jref@jnl{Planet.~Space~Sci.}}   
\def\procspie{\jref@jnl{Proc.~SPIE}}   

\title[Orbits of radial migrators and non-migrators]{Orbits of radial migrators and non-migrators around a spiral arm in N-body simulations}
\author[Grand et al.]
{Robert J.J. Grand \thanks{robert.grand.10@ucl.ac.uk}, Daisuke Kawata, Mark Cropper\\
 Mullard Space Science Laboratory, University College London, Holmbury St. Mary, Dorking, Surrey, RH5 6NT}

\pagerange{\pageref{firstpage}--\pageref{lastpage}} \pubyear{2012}

\pagerange{\pageref{firstpage}--\pageref{lastpage}}\pubyear{2011}
\def\LaTeX{L\kern-.36em\raise.3ex\hbox{a}\kern-.15em
  T\kern-.1667em\lower.7ex\hbox{E}\kern-.125emX}
    
\begin{document}

\label{firstpage}
\maketitle

\begin{abstract}
Recent numerical N-body simulations of spiral galaxies have shown that spiral arms in N-body simulations do not rotate rigidly as expected in classic density wave theory, but instead seem to rotate at a similar speed to the local rotation speed of the stellar disc material. This in turn yields winding, transient and recurrent spiral structure, whose co-rotating nature gives rise to changes in the angular momentum (radial migration) of star particles close to the spiral arm at many radii. From high resolution N-body simulations, we highlight the evolution of strongly migrating star particles (migrators) and star particles that do not migrate (non-migrators) around a spiral arm. We investigate the individual orbit histories of migrators and non-migrators and find that there are several types of migrator and non-migrator, each with unique radial evolution. We find the important quantities that affect the orbital evolution to be the radial and tangential velocity components in combination with the azimuthal distance to the spiral arm at the time the star particle begins to feel tangential force. We contrast each type of orbit to compare how these factors combine for migrators and non-migrators. We find that the positive (negative) migrators sustain a position behind (in front of) the spiral arm, and feel continuous tangential force as long as the spiral arm persists. This is because the positive (negative) migrators are close to the apocentre (pericentre) epicycle phase during their migration, and rotate slower (faster) than the co-rotating spiral arm. On the other hand, non-migrators stay close to the spiral arm, and pass or are passed by the spiral arm one or two times. Although they gain or lose the angular momentum when they are behind or in front of the spiral arm, their net angular momentum change becomes close to zero. We discuss also the long term effects of radial migration on the radial metallicity distribution and radial angular momentum and mass profiles.
\end{abstract}


\section{Introduction}

Radial migration refers to a change in the guiding centre of a star as it orbits the galactic centre. The principal phenomenon that causes radial migration is the interaction with non-axisymmetric structure such as spiral arms, especially around the co-rotation radius \citep[e.g.][]{GLB65,LBK72,SB02}. Unlike other mechanisms such as molecular cloud scattering \citep{SS51,SS53} and diffusion through phase space \citep{BL88,BCP11}, radial migration by spiral arms does not necessarily increase the orbital eccentricity of the stars \citep{LBK72}. 

The phenomenon was first discovered by \citet[]{GLB65} and \citet{LBK72}. For a standing, long-lived density wave pattern, the angular rotation speed of the spiral arms, the pattern speed $\Omega _p$, is constant over the entire radial range of the spiral structure. There is a single radial point in the disc where the spiral arms rotate at the same speed as the disc material. A star near the spiral arm at this radius may therefore be accelerated or decelerated such that the star guiding centre moves away from or towards the galactic centre respectively. This co-rotation radius was found to be the only radius in the disc around which radial migration from spiral arms without heating can occur \citep{LBK72,LA03}. 

\citet{SB02} showed that the radial migration process can have long term effects on the entire stellar disc (which they term radial mixing) if the disc develops successive transient spiral arms. Transient spiral arms are seen in all simulations and their existence represents a departure from classic density wave theory \citep{LS64}, in which the spiral arms are long-lived and standing wave features. Instead, while the reduction of numerical noise as a result of improved resolution \citep[$N>10^6$, as described in][]{Fu11} enables the spiral morphology of galaxies to persist for many rotation periods, simulations repeatedly show that individual spiral arms disappear and reappear on the timescale of a galactic rotation. 

Radial migration produced from transient spiral arms has been highlighted in many recent numerical studies \citep{R08,MF10,Fu11,GKC11,GKC12,MFC11,MFQD12,FB12,RD11,SoS12}, although the precise mechanism of the radial migration process is still debated, which stems mainly from uncertainties in our understanding of the spiral arm nature. For example, \citet{MQ06,MF10,CQ12,CSQ13} explain the transient spiral arms as a superposition of multiple density wave mode patterns that span separate radial ranges that overlap. Because inner patterns rotate faster than outer patterns, they are said to constructively interfere periodically which causes the growth and decay of a spiral arm on the timescale of an interference. On the other hand, other studies \citep[e.g][]{WBS11,GKC11,GKC12,Fu11,FB12} report the spiral arm to be an amplified over-density whose rotation speed matches that of the disc material at all radii. Such spiral arms are naturally winding and transient as shown by \citet{WBS11} who reports transient spiral arms that exhibit a very smoothly decreasing pattern speed as a function of radius. The co-rotating spiral arm is supported by \citet{GKC11,GKC13}, who performed N-body simulations of non-barred galaxies embedded in a static dark matter halo potential and traced the spiral arm peak density directly from the density distribution. This is further backed-up by \citet{RFV13}, who performed high resolution N-body simulations with a live dark matter halo. 

The orbital evolution of particles that radially migrate has to date been discussed in few studies. \citet{RD11} explain that in a spiral disc of multiple co-existing density wave patterns, radial migration occurs only at the co-rotation radius of each pattern. From a sample of star particles chosen from the top $10\%$ of migrators, they interpret radial migration over a large radial range as two successive discrete particle-pattern interactions, where a particle may be transported from the co-rotation radius of one pattern close to the location of the other pattern. In our previous work, \citet{GKC11,GKC12}, we showed that radial migration can occur continuously over a large radial range until the spiral arm disappears. We examined individual N-body star particles and found that the orbital eccentricity was largely conserved. \citet{BSW12} study the orbit evolution of star particles in their high resolution N-body simulation. However, they focus on a random sample of star particles associated with the ``non-steady'' spiral arms in order to link the formation and disruption of spiral arms to the motions of its constituent stars, with less focus on radial migration.   

In this paper, we complement these studies and build upon our own work by running a high resolution simulation of a galactic disc to explore in detail the interaction of star particles with the spiral arm. In particular we focus on star particles that show significant radial migration (migrators), and star particles that show almost no migration (non-migrators) between the birth and death of a spiral arm. We present detailed step-by-step evolution of each group of particle, which reveals several types of migrators as well as non-migrators each with different orbital characteristics, none of which (to our knowledge) have been reported in literature (including our previous works, \citealt{GKC11,GKC12}). Individual orbits are tracked extensively to cover the time before, during and after a single spiral-particle interaction. Our spiral arm peak tracing method \citep{GKC12} enables us to follow the evolution of the particle position with respect to the spiral arm, which is a quantity currently unexplored in the literature. This is an important diagnostic that allows us to identify and explain the properties of the migrating/non-migrating star particles in our Milky Way sized simulation. Therefore, these types of orbits may be observable for the Galaxy in Galactic surveys such as RAVE \citep[e.g.][]{SZ06,Pa12,Pa12b,SFB12,WSB13}, \emph{Gaia} \citep[e.g.][]{LBB08}, Gaia-ESO \citep{GRAB12}, APOGEE \citep[e.g.][]{APM08,BP12}, SEGUE \citep[e.g.][]{YRN09,LBA11}, LAMOST \citep[e.g.][]{CHY12} and 4MOST \citep{dJ12}. 

The main results of this paper come from the detailed analysis of many individual star particle orbits, hence for brevity we show the results from one high resolution simulation. However we briefly discuss the applicability of these results to other simulations of different spiral structure, and discuss the effects of radial migration on the global properties, such as the metallicity and angular momentum distribution.

This paper is organised as follows. In Section 2, we describe the simulation. In Section 3, we describe our particle selection. In Section 4 we analyse the group evolution of the samples in various phase space projections and describe overall macroscopic behaviour. We then examine the orbits of individual particles and categorise several types of migrators and non-migrators in Section 5. The orbital characteristics of each type reveal determining factors, which when combined together distinguish the migrators from the non-migrators. In section 6, we discuss the applicability of the results to other simulations, and briefly show the evolution of the global mass and metallicity distributions as a consequence of the radial migration. In section 7, we summarise our conclusions.

\section{Simulation}

The simulation in this paper is performed with a Tree N-body code \citep{KG03,KOG13}, and simulates a galaxy comprised of a spherical static dark matter halo and a live stellar disc only \citep[for a more detailed description of the simulation set-up, see][]{GKC11}. 

A live dark matter halo can respond to the self-gravitating stellar disc by exchanging angular momentum. This is prominent on long timescales for long-lived non-axisymmetric structures such as a bar \citep{DS00,A02,A12}. However, the effect of the live dark matter halo is expected to be small for transient spiral arms, which justifies the use of a static dark matter halo for our investigation. Furthermore, for practical reasons a live dark matter halo is often modelled with particles more massive than disc particles, which may introduce some scattering and heating that depends on the scale of the mass difference between the particle species. Therefore, in the interest of computational speed and a more controlled experiment, we use a static dark matter halo.

The dark matter halo density profile follows that of \citet{NFW97} with the addition of an exponential truncation term \citep{RA11}:

\begin{eqnarray}
\rho _{\rm dm} = \frac{3 H_0^2}{8 \pi G} \frac{\Omega _0-\Omega_b}{\Omega_0 } \frac{\rho _{\rm c}}{ cx(1+cx)^2}\exp(-x^2) ,
\label{eq1}
\end{eqnarray} 
where $\rho _{\rm c}$ is the characteristic density described by \citet{NFW97}, the concentration parameter, $c = r_{\rm 200}/r_{\rm s}=20$, and  $x= r/r_{\rm 200}$. The truncation term, $\exp(-x^2)$, is needed to ensure that the total mass is convergent. It does not change the dark matter density profile in the inner region of the disc, which is the focus of this paper. The scale length is $r_{\rm s}$, and $r_{\rm 200}$ is the radius inside which the mean density of the dark matter sphere is equal to 200$\rho _{\rm crit}$ (where $\rho _{crit} = 3 H_0^2 / 8 \pi G$; the critical density for closure):

\begin{eqnarray}
r _{200} = 1.63 \times 10^{-2} \left( \frac{M_{200}}{h^{-1} M_{\odot}} \right)^{\frac{1}{3}} h^{-1} \rm kpc,
\end{eqnarray} 
where $M_{200}=1.5 \times 10^{12}$ $\rm M_{\odot}$, $h=H_0/100$ $\rm km$ $\rm s^{-1}$ $\rm Mpc^{-1}$. We assume $\Omega _0 = 0.266$, $\Omega_b = 0.0044$ and $H_0=71$ $\rm km$ $\rm s^{-1}$ $\rm Mpc^{-1}$.

The stellar disc is assumed to follow an exponential surface density profile:

\begin{eqnarray}
\rho _{\rm d,*} = \frac{M_{\rm d,*}}{4 \pi z_{\rm d,*} R_{\rm d,*}^2} {\rm sech}^2 \left(\frac{z}{z_{\rm d,*}}\right) {\rm exp}\left(-\frac{R}{R_{\rm d,*}},\right).
\end{eqnarray} 
where $M_{\rm d,*} = 5 \times 10^{10}$ $\rm M_{\odot}$ is the disc mass, $R_{\rm d,*} = 3.5$ kpc is the disc scale length and $z_{\rm d,*} = 350$ pc is the disc scale height. The simulation presented in this paper has $N=1 \times 10^7$ particles in the stellar disc, so that each star particle is $5000$ $\rm M_{\odot}$. This is sufficient to minimise numerical heating from Poisson noise \citep{Fu11,Se13}. We apply a fixed softening length of $160$ pc (Plummer equivalent softening length of $53$ pc) for star particles with the spline softening suggested by \citet{PM07}.

\section{Particle selection of strong migrators and non-migrators}

\begin{figure*}
\begin{center}

  \subfloat{\includegraphics[scale=0.3]{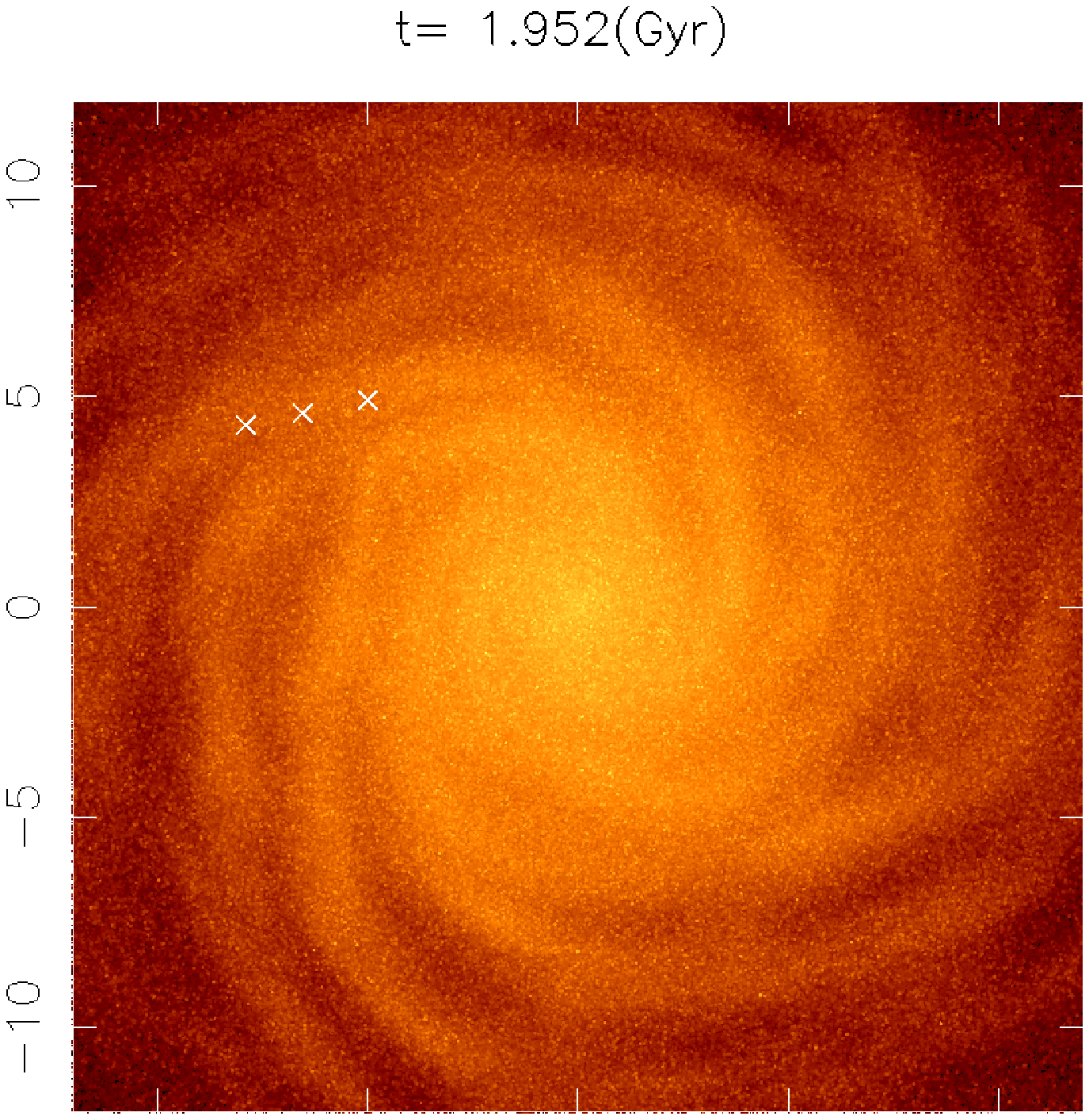}} 
  \subfloat{\includegraphics[scale=0.3] {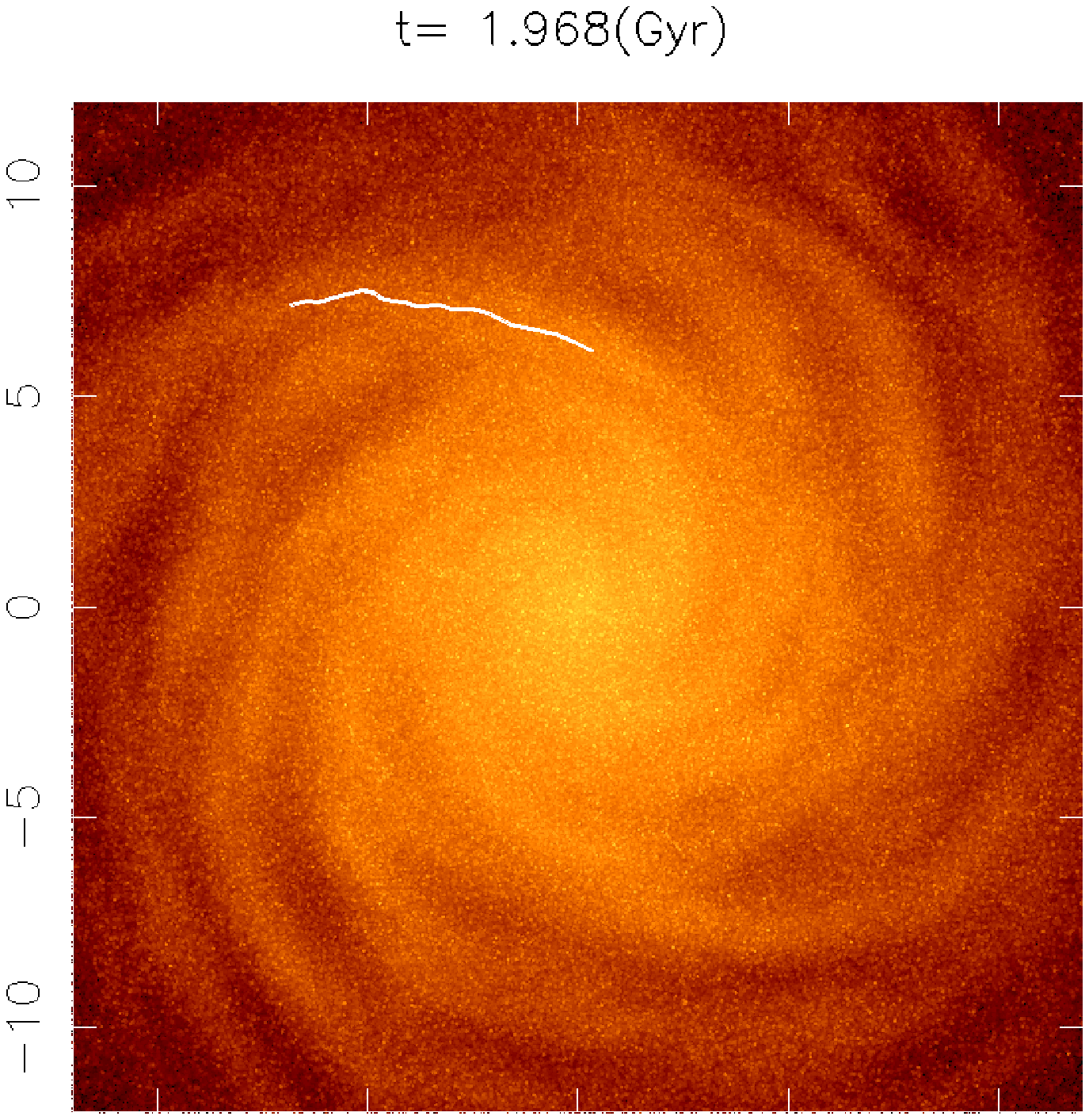}} 
   \subfloat{\includegraphics[scale=0.3] {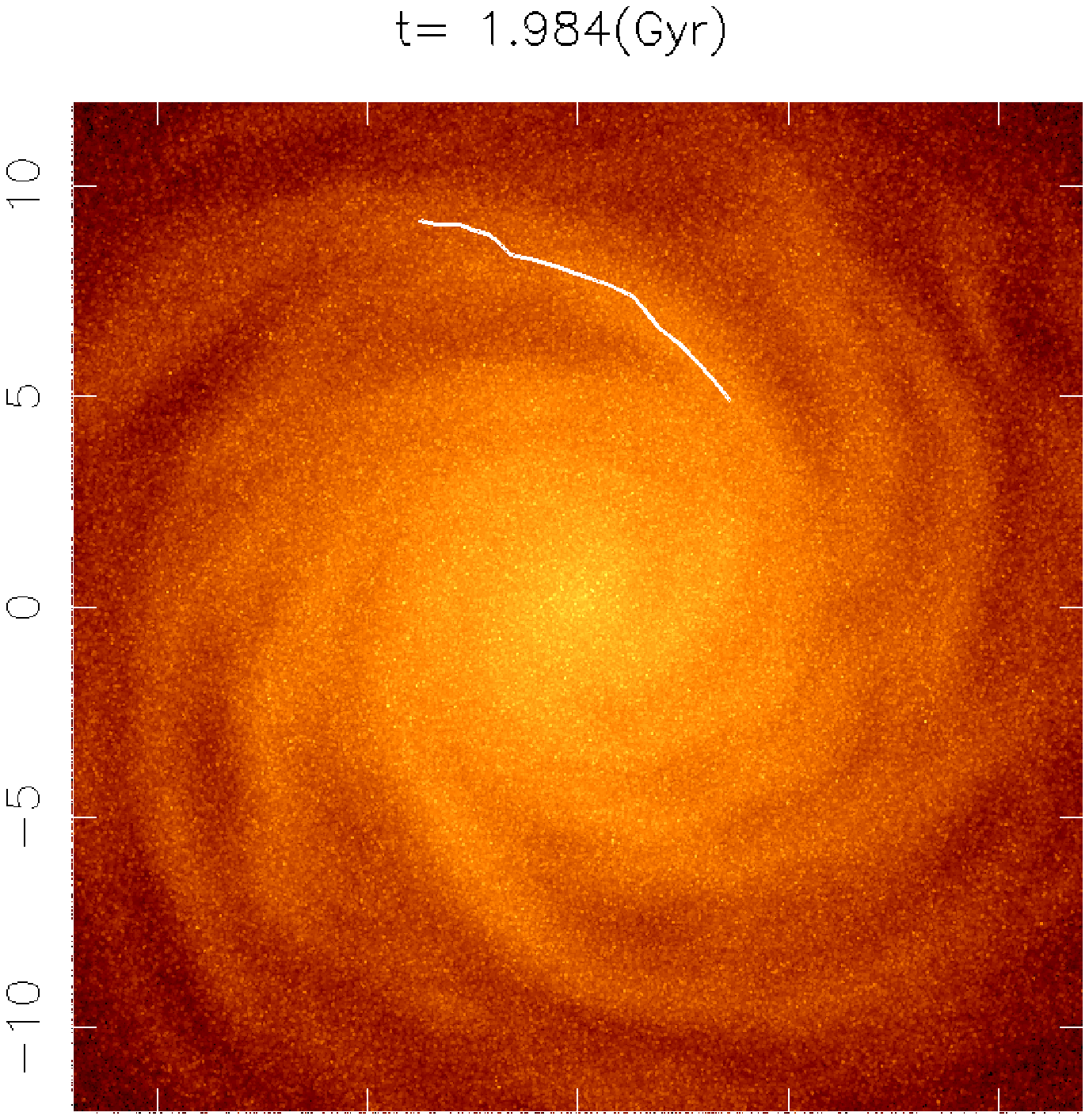}} 
  \subfloat{\includegraphics[scale=0.3] {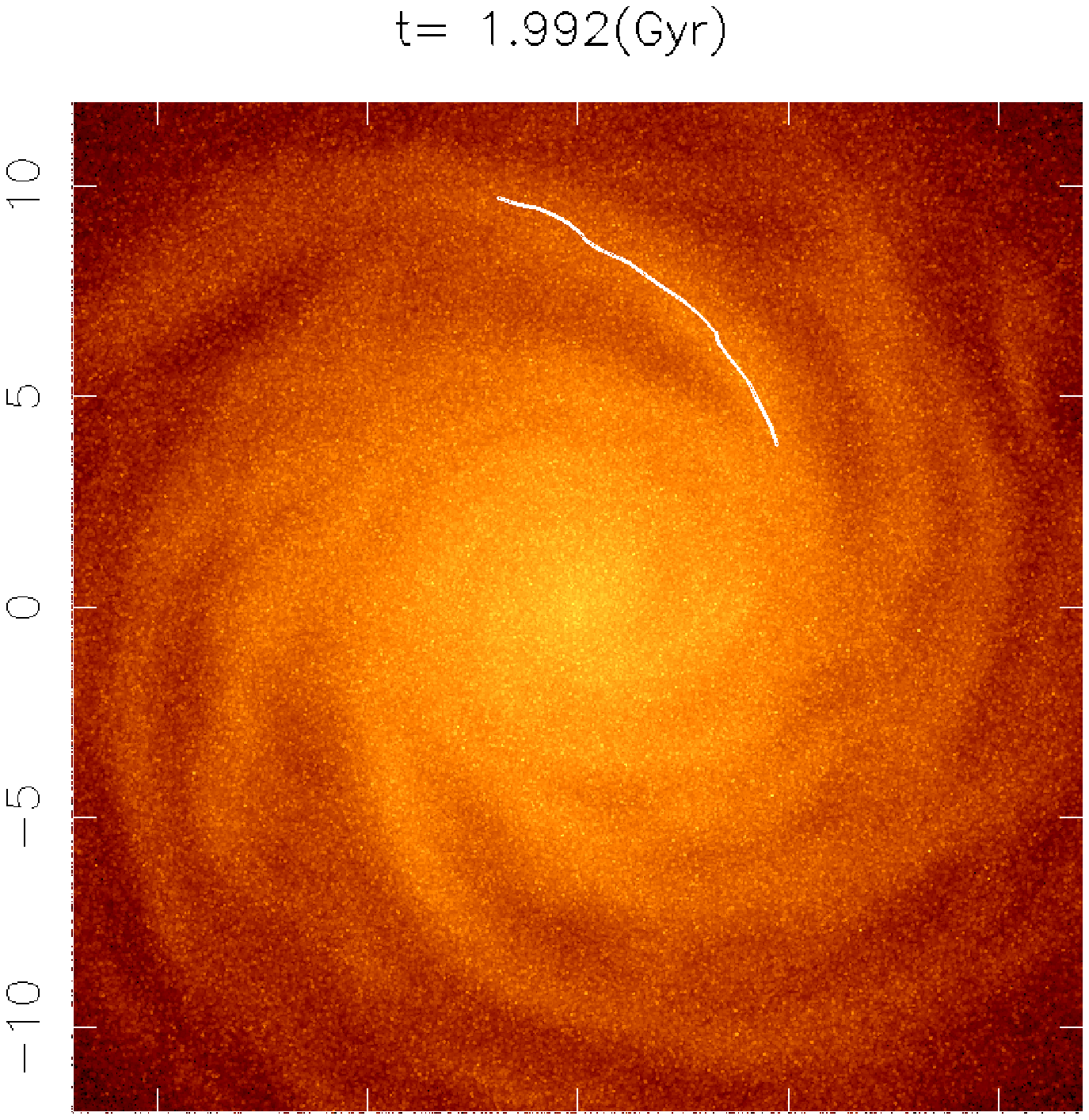}} \\
 \subfloat{\includegraphics[scale=0.3]{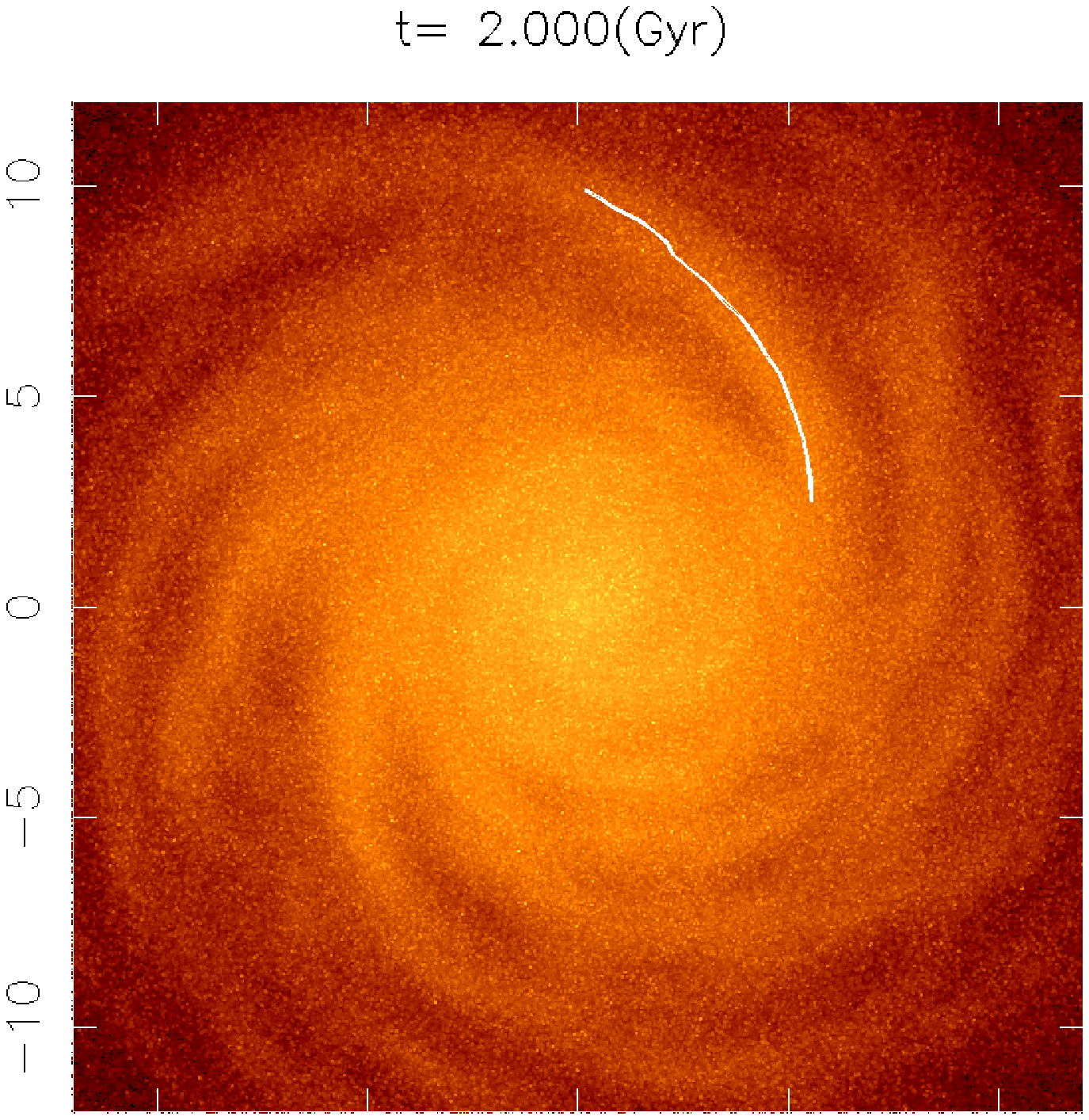}}
   \subfloat{\includegraphics[scale=0.3] {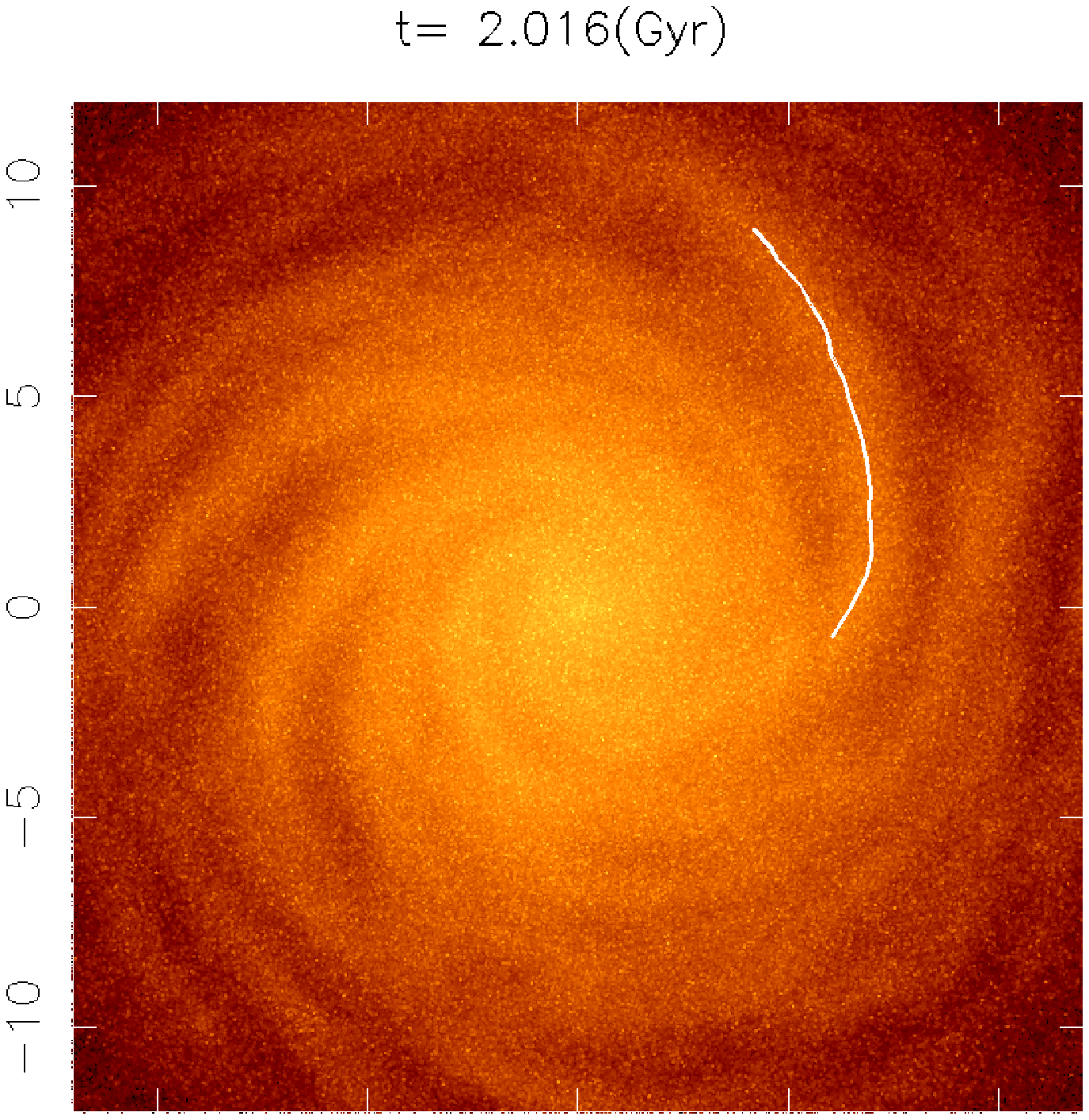}} 
    \subfloat{\includegraphics[scale=0.3] {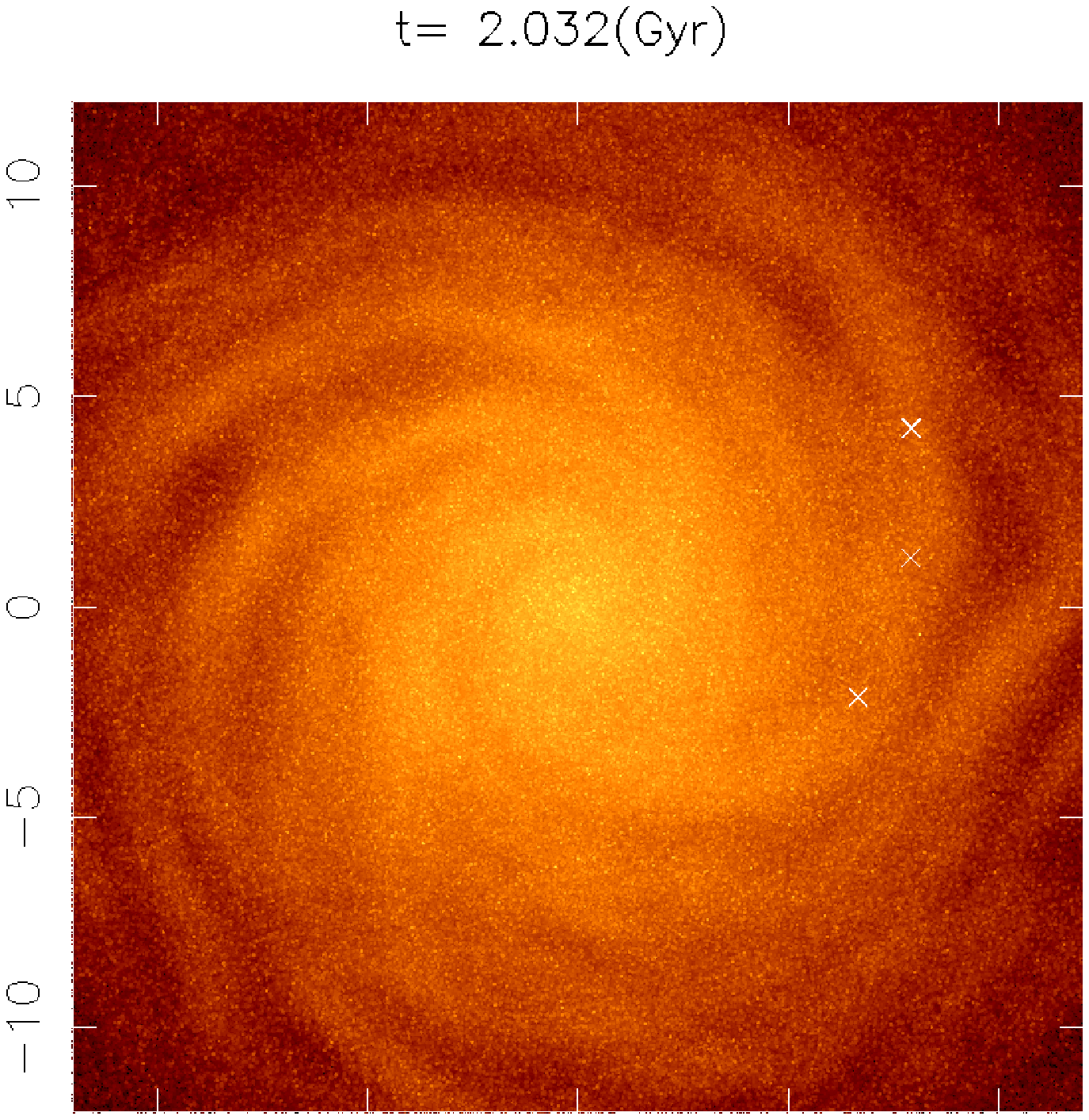}} 
  \subfloat{\includegraphics[scale=0.3] {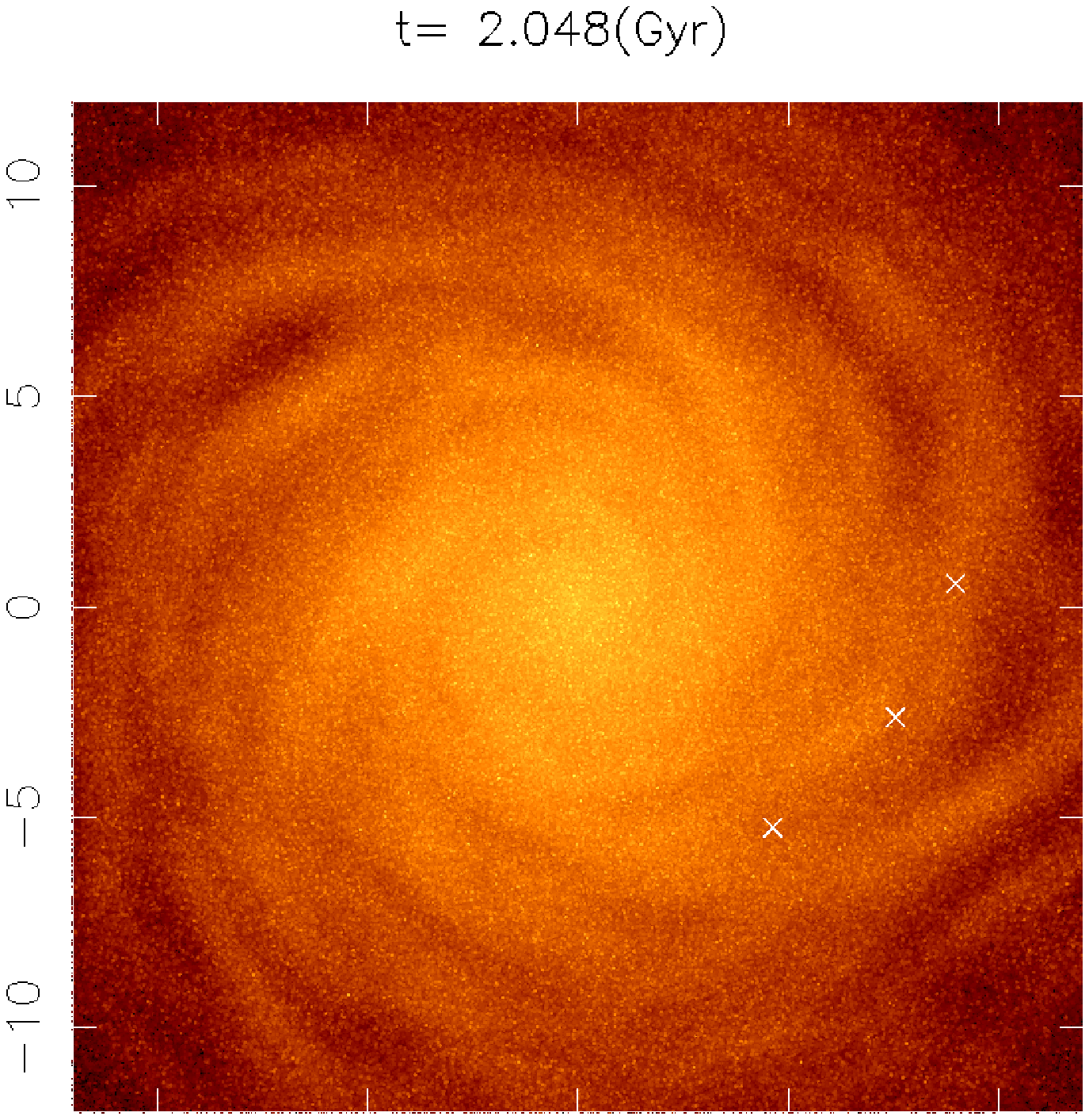}} \\

\caption[]{The face-on density snapshots showing the sequence of the traceable spiral arm. The white line indicates the traced spiral arm of interest. The spiral peak position for $R=7,8$ and $9$ kpc radii at the $t=2.0$ Gyr snapshot are rotated with the corresponding circular velocity and marked as anchors (white crosses) on the $t=1.952, 2.032$ and $2.048$ Gyr snapshots, to guide the eye to the forming and disrupting stages of the spiral arm respectively.}
\label{tracea1}
\end{center}
\end{figure*} 

In this study, we focus on single spiral arm-star particle interactions on the timescale of the spiral arm lifetime, which we find to be $\sim 100$ Myr. We scrutinize the evolution of a sample of star particles in order to obtain detailed information on these interactions.  

\begin{figure}
\centering
\includegraphics[scale=0.45]{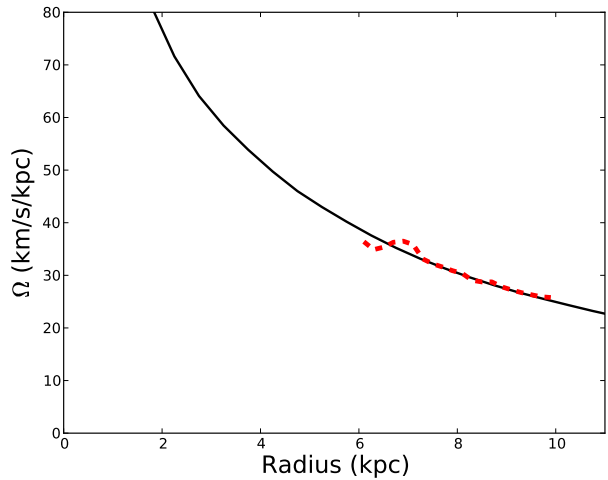}
\caption[]
{The pattern speed (red dashed line) of the traced spiral arm highlighted above. The angular speed of the stellar disc is also plotted (solid black line). The pattern speed matches the angular speed of the disc material well.}
\label{omegap}
\end{figure}

The first step is to identify coherent spiral arms for which we can reliably trace the spiral arm peak position. A reliable trace is defined as a radial range for which each radial point shows a smooth single density peak in azimuth. A snapshot in which the spiral arm exhibits a double peak structure anywhere within the defined radial range for tracing is rejected. Double peak features are usually associated with the formation and destruction stages of spiral arm evolution, and are not suitable for an unambiguous trace \citep[see][for more details]{GKC12}. Fig. \ref{tracea1} shows the face-on density snapshots of our selected  spiral arm.  There is a time window over which the spiral arm can be reliably traced, and a wider time window which will be used to examine the prior and subsequent star particle behaviour. We could trace the spiral arm reliably at the radii between 6 and 10 kpc in the traceable time window spanning $t = 1.968$ to $2.024$ Gyr (highlighted with the white line in Fig. \ref{tracea1}). Outside of this time window, we extrapolate the spiral arm position by rotating the $R=7,8$ and $9$ kpc peak positions of the $t=2.0$ Gyr snapshot with the circular velocity (anchors marked with white crosses) to guide the eye to the spiral arm when it forms and disrupts. The pattern speed is calculated by simple subtraction of the peak line between snapshots. Fig. \ref{omegap} shows the radial profile of the time-averaged pattern speed (dashed red line) and the angular rotational velocity of the disc (solid black line). The latter is calculated from the radial force averaged over azimuth for each radius. Fig. \ref{omegap} confirms that the spiral arm is co-rotating with the rotational velocity of the disc \citep[as shown in][]{GKC11,RFV13}. For the rest of the analysis, we focus on particle interactions with this spiral arm. 

We choose a snapshot at a time when the spiral arm is fully formed and adopt $t=2.0$ Gyr  (Fig. \ref{tracea1}) as the time of particle selection. From the traced spiral arm at this time, we define a region within a range of $4$ kpc either side of the density peak of the spiral arm in the azimuthal direction. The region is bounded by a radial range of $6$ - $10$ kpc and a vertical height of $|z| < 0.1$ kpc. The sample is constituted of particles found inside this spatially defined region at $t=2.0$ Gyr. This probes a wide range in radius and azimuthal position with respect to the spiral arm, and restricts the majority of particles to be in the plane of the disc because more migration takes place in the plane of the disc. Note that over time, the vertical oscillations cause some particles to move to heights that exceed the selection cut of $|z|< 0.1$ kpc. We found that about $15 \%$ of the star particles selected reach $z_{max}>0.35$ kpc (one initial galactic scale height). However, we find that there is no significant difference between the trends discussed in this paper for particles of different maximum heights.

\begin{figure}
\centering
\subfloat{\includegraphics[scale=0.61]{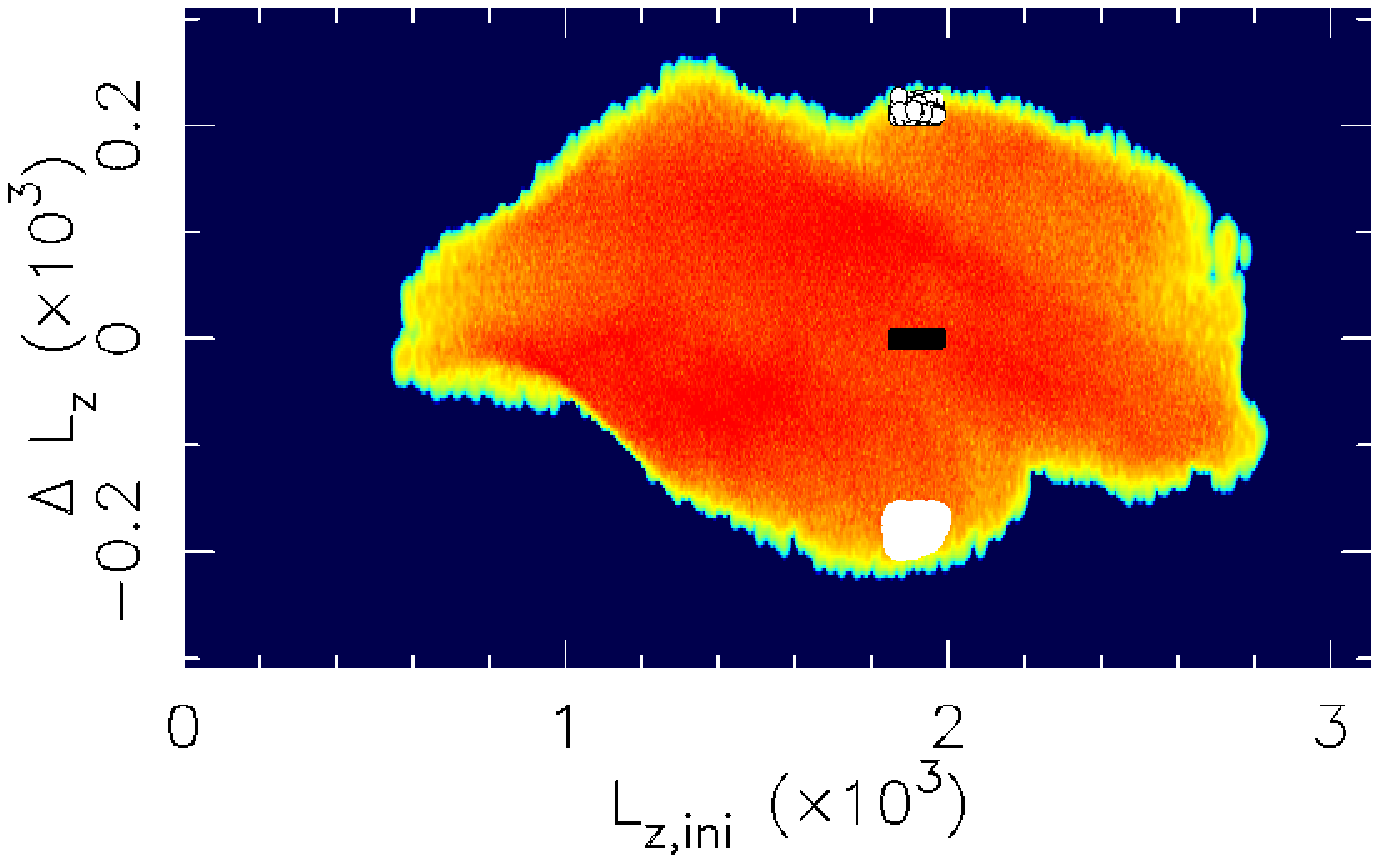}}
\caption[]
{The change in angular momentum of the sample of particles over the time window $T_{fin} - T_{ini}$, as a function of their initial angular momentum. Over-plotted are the strong positive migrators (white with black outline symbols), strong negative migrators (white symbols) and non-migrator particles (black symbols) selected. The units are $\rm kpc$ $\rm km$ $\rm s^{-1}$.}
\label{deltaL}
\end{figure}

The time at which the star particles are selected is defined as the central time step, $T_{c}=2$ Gyr, and a time window is then defined as $\Delta T = T_{fin} - T_{ini}$, where $T_{ini} = T_c - 48$ Myr and $T_{fin} = T_c + 48$ Myr. This time window spans the $\sim 100$ Myr lifetime of the spiral arm. Note that this time window is longer than the time window for which we could trace the spiral arm. However, as seen in Fig. \ref{tracea1}, the spiral arm begins to form at around $t=1.952$ Gyr and disrupts at around $t=2.048$ Gyr. The motion of nearby star particles can be affected at these times. In fact, we will demonstrate in Section. 5 that the motion of some star particles can be affected as early as $t\sim 1.9$ Gyr and last until $t\sim 2.2$ Gyr. Hence this time window is set by convenience and not a strict definition of the formation and destruction time. The star particle sample is then plotted in the $L_{z,ini} - \Delta L_z$ plane, where $L_{z,ini}$ is the $z-$component of angular momentum of the star particles at the beginning of the time window, $T_{ini}$, and $\Delta L_z$ is the change in angular momentum from the initial time, $T_{ini}$, to the final time, $T_{fin}$. This is plotted in Fig. \ref{deltaL}. From Fig. \ref{deltaL}, we see that there is a wide range of initial angular momentum values over which the angular momentum i.e. the guiding centre, is changed, which is consistent with our previous studies \citep{GKC11,GKC12}. 

From the star particle distribution in Fig. \ref{deltaL}, we select a sample in the range  $ 1.86\times10^3 < L_{z,ini} < 1.97\times10^3 $ $\rm kpc$ $\rm km$ $\rm s^{-1}$ (particle samples at other $L_{z,ini}$ exhibit similar behaviour, so we focus on one sample), which corresponds to a guiding centre radius of about 8 kpc. The sample is further cut into subgroups: strong migrators (both negative and positive) and non-migrators. The strong positive (negative) migrators consist of those particles that have the largest positive (negative) $\Delta L_z$, and the non-migrators are those that have the lowest changes in angular momentum. These star particles are selected such that there are $\sim 200-300$ particles in each star particle group, and are highlighted in Fig. \ref{deltaL}.

\begin{figure*}
\begin{center}
  \includegraphics[scale=0.8] {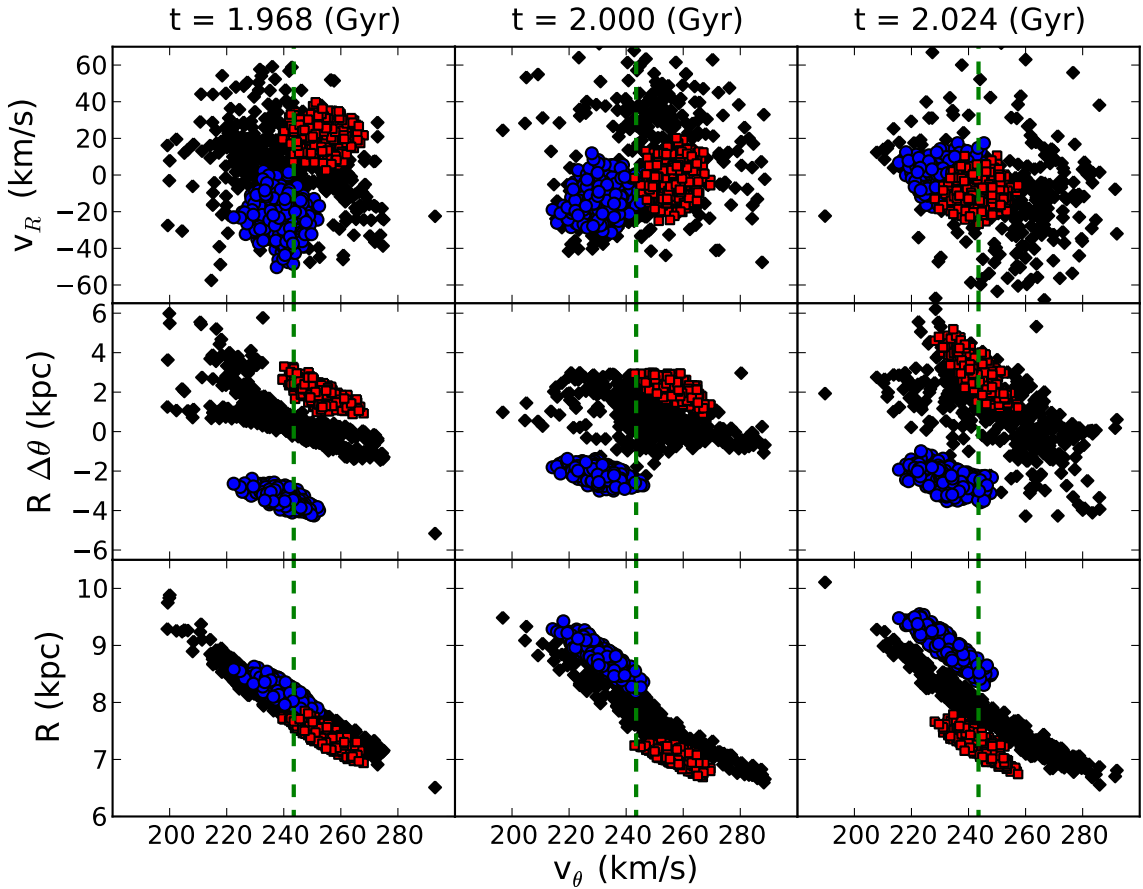}
\caption[]{\emph{Top row:} The negative migrators (red squares), positive migrators (blue circles) and non-migrators (black diamonds) of the sample in $v_{\theta} - v_R$ space. \emph{Middle row:} Shows the azimuthal distance between the star particles and the spiral arm peak position, $R\Delta \theta$, as a function of azimuthal velocity, $v_{\theta}$. \emph{Bottom row:} Plots the radius of the same star particles with $v_{\theta}$. Each column shows these projections at three time epochs, increasing from left to right. The circular velocity at the $8$ kpc radius, $v_c = 243.5$ $\rm km$ $\rm s^{-1}$, is marked in each panel by the vertical dashed green line.}
\label{phase67}
\end{center}
\end{figure*}

\section{Evolution of sample in phase space}

\begin{figure}
\begin{center}
   \subfloat{\includegraphics[scale=2.8] {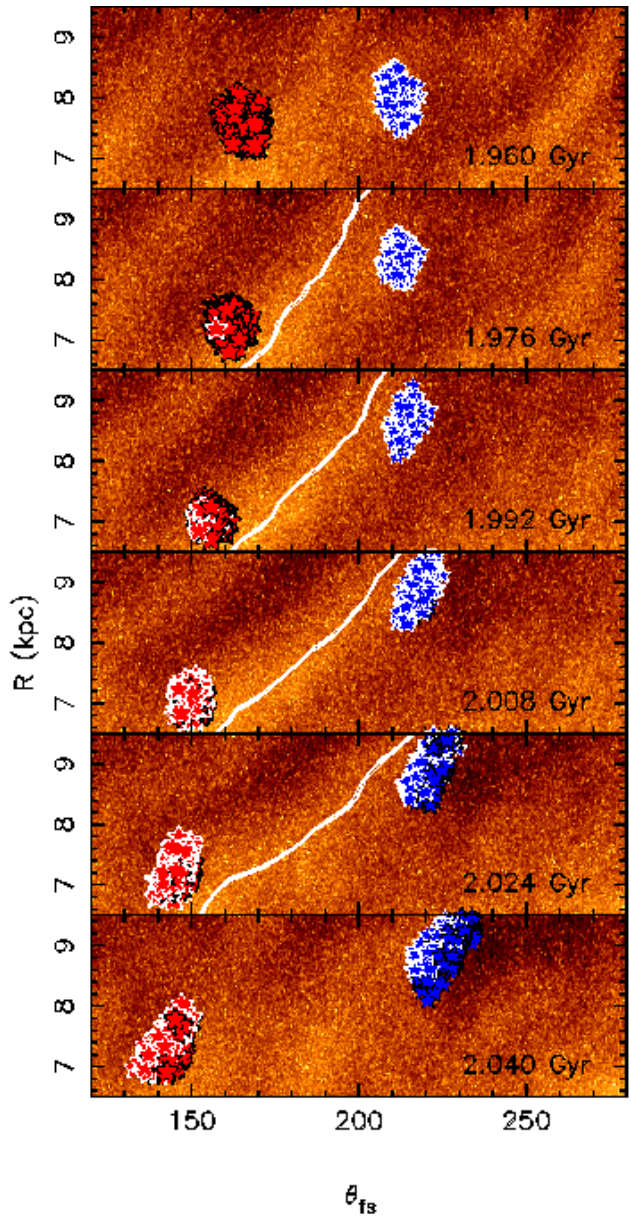}} 
\caption[]{Time sequence of a close up of the density map in the $R-\theta _{fs}$ plane. The rotation is from right to left in a rotating frame that co-rotates with the circular velocity at $R=8$ kpc. Positive migrators are marked as blue stars, and negative migrators are marked as red stars. The radial velocity direction is indicated by the white and black outline of the symbols, which represent outward and inward moving radial velocities respectively. The white line (present in some panels) highlights the peak position of spiral arm at each radius.}
\label{pdot67}
\end{center}
\end{figure} 

\begin{figure}
\begin{center}
   \subfloat{\includegraphics[scale=2.8] {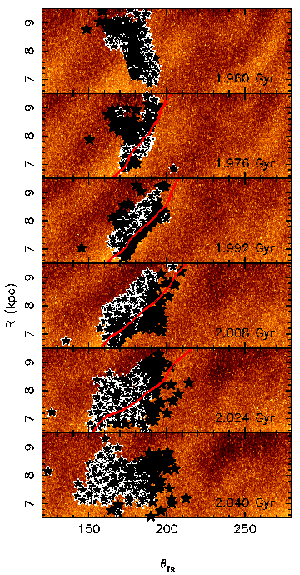}} 
\caption[]{The same as Fig. \ref{pdot67} but showing non-migrator particles. The red line indicates the spiral arm peak position.}
\label{pdotn67}
\end{center}
\end{figure}

We plot each particle in the sample in various projections in phase space, and highlight the positive migrator (blue circles), negative migrator (red squares) and non-migrator (black diamonds) groups in Fig. \ref{phase67}. Each column corresponds to a different time: the left column shows the sample at the earliest time that the spiral arm could be reliably traced \citep[following criteria in][]{GKC12}, the middle column is the time of selection, $T_c$, and the right column is the latest time that the spiral arm could be reliably traced. 

The top row of Fig. \ref{phase67} shows the sample plotted in the radial velocity, $v_R$ - azimuthal velocity, $v_{\theta}$, plane. Positive radial velocity, $v_R > 0$, is in the direction of the galactic centre. The circular velocity at $R=8$ kpc is $v_c(R=8) = 243.5$ km s$^{-1}$, which is marked by the dashed green lines in Fig. \ref{phase67}. The positive and negative migrators appear to occupy separate regions of velocity space, whereas the non-migrators (black diamonds) are more evenly distributed and overlap the migrator groups. The positive migrators have outward facing velocities ($v_R < 0$) during their outward migration. It is interesting to see that their azimuthal velocity tends to be slower than the circular velocity at $R=8$ kpc. The opposite is applied to the negative migrators, which move inward ($v_R > 0$) and rotate faster than the circular velocity at $R=8$ kpc.

The azimuthal distance of a star particle with respect to the spiral arm, $R\Delta\theta$, is defined as the length of an arc that joins the star particle azimuth position to the spiral arm peak azimuth position at that radius. The second row of Fig. \ref{phase67} shows this quantity as a function of azimuthal velocity for the sample. In this plane, each group of particles (including non-migrators) is very clearly separated. The positive migrators always stay behind the spiral arm ($R\Delta\theta < 0$), and the negative migrators always stay in front of the spiral arm ($R\Delta\theta > 0$), throughout the traceable spiral arm lifetime. The non-migrators are clustered around the spiral arm. At the $t = 1.968$ Gyr (left panel), the arrangement of each particle group is highly ordered. The non-migrator group in particular is spread over a large range of azimuthal velocity, which appears tightly correlated with the azimuthal distance between star particle and spiral arm. For example, at $t=1.968$ Gyr, at $v_{\theta}=220$ $\rm km$ $\rm s^{-1}$ the positive migrators have a negative $R\Delta\theta$, i.e. they are behind the spiral arm, while the non-migrators have a positive $R\Delta\theta$, i.e. they are in the front of the spiral arm. Conversely, at $v_{\theta} \sim 265$ $\rm km$ $\rm s^{-1}$, the negative migrators have a positive $R\Delta\theta$, i.e. they are in the front of the spiral arm, while the non-migrators have a negative $R\Delta\theta$, i.e. they are behind the spiral arm. In Section. 5.3, we demonstrate that non-migrators must pass or be passed by the spiral arm at some point during the spiral arm lifetime. Roughly speaking, the star particle will cross the spiral arm if: $| R\Delta\theta | < | \int_{t_0}^{t_1} v_{\theta ,sub} dt |$, where $t_1-t_0 < \Delta T$, and $v_{\theta ,sub} = v_{\theta} - v_c$, where $v_{\theta}$ is the azimuthal velocity of the star particle and $v_c$ is the circular velocity at the particle radius. Although these star particles were selected at $t=2.0$ Gyr, they are more ordered at $t=1.968$ Gyr. This indicates that the $R\Delta\theta - v_{\theta}$ phase space can be diagnostic at the early stages of spiral arm formation (middle-left panel of Fig. \ref{phase67}) in pre-determining whether a star particle will be a migrator or a non-migrator. At later times, the groups become less clearly separated, but still maintain the trend.

The bottom row of Fig. \ref{phase67} shows the sample in the $R - v_{\theta}$ plane. At $t = 1.968$ Gyr (left panel), migrator and non-migrator particle groups occupy the same region of this space, and become more separated at the later times (middle and right panels of Fig. \ref{phase67}) owing to the migration taking place. The distribution of non-migrators in this plane highlights the epicyclic motion of the star particles. Star particles that are at a radius greater than the guiding centre of the sample, $R > R_g \sim 8$ kpc, possess azimuthal velocities lower than the circular velocity at the guiding centre, $v_{\theta} < v_c(R_g)$, whereas star particles at a radius smaller than the guiding centre possess azimuthal velocities larger than the circular velocity at the guiding centre. Positive migrators obviously move toward larger radii, and are outside of their guiding centre owing to their relatively low rotation velocity with respect to the circular velocity at that radius. In other words, the positive migrators are always close to the apocentre phase during their migration, and the negative migrators are always close to the pericentre phase.

Fig. \ref{pdot67} shows the evolution of both positive migrators (blue stars) and negative migrators (red stars) plotted on the face-on density maps of the disc in the $R-\theta_{fs}$ plane. The evolution is shown in a rotating frame that co-rotates with the circular velocity at $R=8$ kpc. The $\theta$ coordinate of all star particles has been subtracted by an amount corresponding to the rotating frame such that the spiral arm and sample particles remain within the 120-280 degree azimuth window i.e. $\theta _{fs} = \theta _{true} - \Omega _{fr} \Delta t$, where $\Delta t = T_{ini} - t$. Here $\Omega_{fr}$ is the angular rotation speed of the frame. The direction of motion is from right to left. Each migrator particle is outlined in white (black) to indicate the outward, $v_R<0$ (inward, $v_R>0$) direction of the radial velocity vector, in order to indicate the epicycle phase i.e. $v_R<0$ means that the star particle is moving from pericentre to apocentre, while $v_R>0$ indicates that the star particle is moving from apocentre to pericentre. Both groups exhibit a range of radial velocities at each snapshot, which indicates there is some spread in the epicycle phase within the groups. For the positive migrators, particles closer to the spiral arm at $t=1.928$ and $2.024$ Gyr have negative radial velocities (approaching apocentre) and particles further from the spiral arm have positive radial velocities (moving away from apocentre). The opposite trend is seen in the negative migrator group. Despite the different epicycle phases of these migrator groups, all star particles in the positive and negative migrator groups radially migrate eventually, as we show below.

Fig. \ref{pdotn67} shows the evolution of the non-migrators in the $R-\theta_{fs}$ density plane in the rotating frame described above for Fig. \ref{pdot67}. The symbols are outlined in black and white corresponding to inward and outward radial velocity unit vectors respectively. Most of the non-migrators are clustered close around the spiral arm (peak position marked in red), and appear spatially separated according to the direction of radial motion. For example, at $t= 1.992$ and $2.008$ Gyr, most star particles behind (in front of) the spiral arm are moving towards pericentre (apocentre). This is a clear contrast from the migrators. At $t=1.976$ Gyr, the positive migrators behind the spiral arm are moving toward apocentre, i.e. outward, while the negative migrators in the front of the spiral arm are moving toward pericentre, i.e. inward.

\section{Individual particle orbits}

In this section, we analyse the orbits of the positive, negative and non-migrator particles of the sample individually. We took $\sim 100$ random samples of each star particle group, and followed the evolution of each orbit individually. We scrutinised the orbits over the course of the particle-spiral arm interaction, and categorise several types of migrators and non-migrators. Below we show an example of each type and outline their defining features. We refer to positive migrators with a suffix `g' because they gain angular momentum, negative migrators with a suffix `l' because they lose angular momentum and non-migrators with a suffix `n'. In Table. \ref{tab1}, we list the orbital properties of each orbital type mentioned below.

\subsection{Orbits of positive migrators}

\begin{figure*}
\begin{tabular} {l r} \hspace{-2.0mm}
 \includegraphics[scale=2.5] {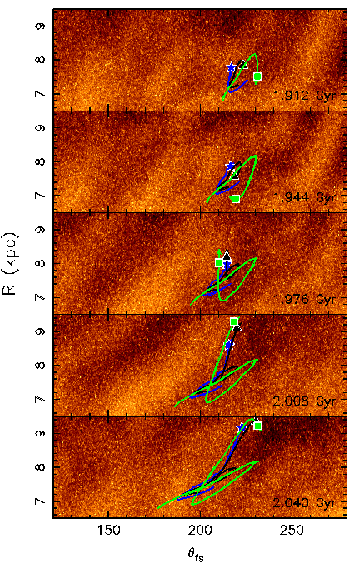} &  \includegraphics[scale=2.4] {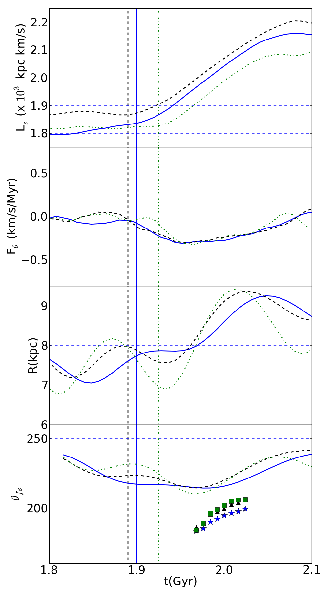} \\
\end{tabular}
\caption{The evolution of three types of positive migrator. \emph{Left panels}: The time evolution of the particles in the close up $R-\theta_{fs}$ density map in a rotating frame equal to the circular velocity at $R=8$ kpc. Rotation is from right to left. The symbols depict the current position of each star particle, and the lines show the history of each star particle orbit relative to the spiral arm (see text for more details). \emph{Top-right panel:} The evolution of the angular momentum of the star particles. \emph{Second-right panel}: The evolution of the tangential force per unit mass, $F_{\theta}$, acting upon the star particles. \emph{Third-right panel}: The radial evolution of the star particles. \emph{Bottom-right panel}: The azimuthal angle of the star particles in the rotating frame of the left panels, $\theta_{fs}$, (lines) and spiral arm azimuthal angle at the particle radius (symbols). The latter can only be calculated in the traceable spiral arm time window defined in Section. 3. In all right-hand panels, the time at which each particle first feels the tangential force is indicated by a vertical line of corresponding line style.}
\label{L9g}
\end{figure*}

Fig. \ref{L9g} shows an example of three different types of positive migrator. In the left panels, we show a selection of snapshots from the time sequence evolution of these particles in the same rotating frame adopted in Fig. \ref{pdot67}. The symbol in each snapshot indicates the position of the particle at the time given in the bottom-right corner of the panel, and the lines show the history of the orbit in the co-rotating frame. Because the spiral arm rotates with the circular velocity (Fig. \ref{omegap}), the spiral arm in the left panels of Fig. \ref{L9g} moves relative to the rotating frame in time: for $R<8$ kpc, the spiral arm moves to lower $\theta_{fs}$, whereas for $R>8$ kpc the spiral arm moves to higher $\theta_{fs}$. We remove the relative motion between the spiral arm and star particle orbit histories by making a further adjustment to the orbit history. We calculate the difference between the velocity of the rotating frame and the circular velocity at the radius of the line point. At each subsequent time step after the line point appears the position of the line point is shifted by the amount corresponding to this velocity difference. The purpose of this adjustment is to give an idea of where the past particle positions were \emph{relative to the spiral arm} at previous times, although it is not exact because it is impossible to place the past orbit around the dynamically changing spiral arm.

The 1st- and 2nd-right panels of Fig. \ref{L9g} shows the evolution of particle angular momentum, $L_z$, and the tangential force per unit mass, $F_{\theta}$, where the positive direction is opposite to the direction of rotation, i.e. from left to right in the left panels of Fig. \ref{L9g}. Blue solid, black dashed and green dot-dashed lines correspond to the blue, black and green lines in the left panels of Fig. \ref{L9g}. 

We define radial migration to be a change in angular momentum over time, such as the change seen in the top-right panel of Fig. \ref{L9g}, e.g. the black dashed line from $t \sim 1.89$ to $2.08$ Gyr when $L_z$ increases. Note that the tangential force is always negative during the increase in $L_z$. The magnitude of the tangential force indicates the rate of change of angular momentum, which allows us to see more clearly when and where the star particles migrate in the left panels of Fig. \ref{L9g}. The third-right and bottom-right panels of Fig. \ref{L9g} show the evolution of the particle radius, $R$, and azimuth angle in the rotating frame, $\theta _{fs}$, respectively. In the bottom panel, the blue star, black triangle and green square indicate the azimuth angle of the spiral arm that we identified in Fig. \ref{tracea1} at the star particle radius at the corresponding time. Note that we could trace the spiral arm for only part of the period when the spiral arm is clearly seen. However, this demonstrates that the spiral arm affects the orbit of the star particles well before the arm is clearly seen and even after it begins to disrupt. At least in this short period when we can clearly trace the spiral arm, we can show the particle position with respect to the spiral arm as seen in the bottom panel of Fig. \ref{L9g}. The evolution of each quantity shown in the right panels of Fig. \ref{L9g} will determine the type of each positive migrator. 

\subsubsection{Type 1g positive migrator}

The first type (Type 1g) of positive migrator (blue star and solid blue line in the left panels and blue solid line in the right panels of Fig. \ref{L9g}) is quite close to the spiral arm when it begins to feel a negative tangential force at around $t \sim 1.9$ Gyr (2nd-right panel of Fig. \ref{L9g}). The top left panel of Fig. \ref{L9g} shows that the spiral arm begins to build up at $t\sim 1.9$ Gyr. It appears that the density enhancements around $\theta_{fs} = 180$ and $\theta_{fs} = 220$ at $R\sim 8$ kpc at $t=1.912$ Gyr merge and form the single spiral arm around $t\sim 1.976$ Gyr. At $t=1.912$ Gyr the blue star is located behind the density enhancement at $\theta_{fs} = 220$, and therefore the direction of tangential force is negative ($F_{\theta} < 0$) and the star particle is accelerated. 

At $t\sim 1.9$ Gyr, the particle is approaching the apocentre phase of orbit (3rd-right panel of Fig. \ref{L9g}). However, as a result of strong negative tangential force at this phase (2nd-right panel of Fig. \ref{L9g}), the radius of the star particle does not decrease again according to normal epicycle motion. This is because of the competition between mainly the radial gravitational force and the increase in centrifugal force caused by the gain in angular momentum. In this case, the radial force is balanced by the increased centrifugal force. As a result, the star particle pauses at a radius of $R \sim 7.9$ kpc for $\sim 20-30$ Myr at around $t \sim 1.94$ Gyr, then increases again once the angular momentum has increased such that the centrifugal force is large enough to overcome the radial gravitational force. 

Note that irrespective of the evolution of radius, the increase in angular momentum is sustained as the particle continues to be accelerated by the negative tangential force, because the particle is always behind the spiral arm. The bottom-right panel of Fig. \ref{L9g} shows that the azimuth angle of the particle is always larger than that of the spiral arm (star particle is behind the spiral arm) during the epoch at which the spiral arm is clearly traced. The strong migrators are able to stay on one side of the spiral arm, because the spiral arm co-rotates with the disc material \citep{GKC11}. As a result, positive migrators maintain their position behind the spiral arm and continue to be accelerated and migrate along the spiral arm. At around $t\sim2.08$ Gyr when the spiral arm is disrupted, the particle stops gaining angular momentum and the star particle resumes epicycle motion.  

\subsubsection{Type 2g positive migrator}

The second type (Type 2g) of positive migrator (black triangle and solid black line in the left panels and black dashed line in the right panels of Fig. \ref{L9g}) begins to feel a negative tangential force at about the same time as the Type 1g positive migrator mentioned above ($t \sim 1.89$ Gyr), and is accelerated. Again, there is a competition between the radial gravitational force and the increase in angular momentum. However, at this time the particle has just passed the apocentre phase of the epicycle motion (3rd-right panel of Fig. \ref{L9g}) when the angular momentum begins to increase. Therefore, the star particle begins to move inward for a while, until the angular momentum increases sufficiently to overcome the gravitational force. As a result, the amplitude of the epicycle motion is shortened. This shortening in amplitude is shown by the changed pericentre radii between $t\sim1.83$ and $1.93$ Gyr in the 3rd-right panel of Fig. \ref{L9g}. The pericentre radius is larger at the later time owing to the increase in guiding centre. Note that although the radial evolution of the orbit looks different to that of the Type 1g positive migrator shown above, the angular momentum steadily increases irrespective of their radial evolution because the star particle is always located behind the spiral arm and accelerated (bottom-right panel of Fig. \ref{L9g}). 

This is the most common type of positive migrator in this sample. The shortened epicycle motion present in the radial evolution of the Type 2g positive migrator is also reported in \citet{RD11}. The strongest migrators in their simulation are shown to exhibit several shortened epicycle motions, which they interpret as effects of co-rotation resonances of two spiral waves: one inner, faster rotating spiral pattern and one outer, slower rotating spiral pattern. For example, in the top panels of Fig. 11 of \citet{RD11}, the orbit of the migrator shows $\sim 6$ epicycle motions from $t\sim 5.4$ to $5.9$ Gyr. In our study, we focus on 1-2 epicycle periods which corresponds to the lifetime of the spiral arm in our simulation. The Type 2g migrator demonstrates that during the shortened epicycle motion the guiding centre of the star particle continuously increases owing to the angular momentum increase at every radius. Therefore, we think that it is difficult to attribute 6 shortened epicycle motions seen in \citet{RD11} to only two co-rotation resonances. Instead, we think that the spiral arm feature co-rotates with the disc material at every radius \citep{GKC11}, and induces a continuous gain in angular momentum of migrators as long as the feature persists. We suspect that their spiral arm is short-lived, and that the 6 epicycle motions seen over $\sim 0.5$ Gyr are affected by several transient spiral arm features that co-rotate and accelerate star particles behind the spiral arm at all radii.

\begin{figure*}
\begin{tabular} {l r} \hspace{-2.0mm}
 \includegraphics[scale=2.5] {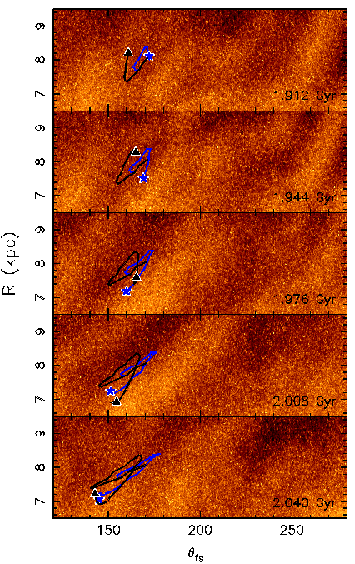} &  \includegraphics[scale=2.4] {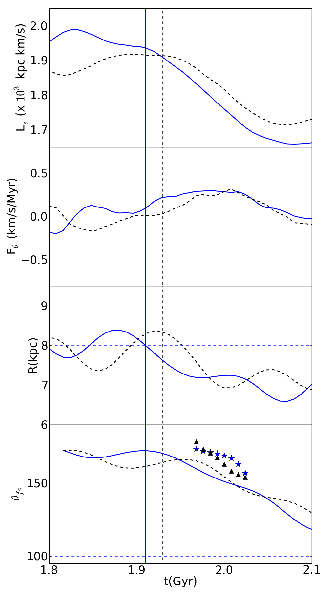} \\
\end{tabular}
\caption{The same as Fig. \ref{L9g} but for types of negative migrator.}
\label{L9l}
\end{figure*}

\subsubsection{Type 3g positive migrators}

The third type (Type 3g) of positive migrator (green square and green solid line in the left panels and green dot-dashed line in the right panels of Fig. \ref{L9g}) originates much farther from the spiral arm as it begins to build up at around $t\sim 1.9$ Gyr (top-left panel of Fig. \ref{L9g}), and consequently begins to feel a negative tangential force at the later time of $t\sim 1.93$ Gyr with respect to Types 1g and 2g. At this time, the star particle approaches the pericentre phase of the epicycle (3rd-right panel of Fig. \ref{L9g}), and therefore it moves closer to the spiral arm at $t\sim 1.976$ Gyr in the left panels and bottom-right panel of Fig. \ref{L9g}. In this case, both the outward epicycle motion and the angular momentum increase facilitate the outward motion of the star particle. As a result, the particle radius increases very rapidly (third-right panel of Fig. \ref{L9g}). 

Note that the star particle reaches the apocentre at $t\sim 2.02$ Gyr, well before the angular momentum gain ends at $t\sim 2.06$ Gyr. This means that around $t\sim 2.02$ Gyr, the radial gravitational force becomes greater than the enhanced centrifugal force provided by the boost in angular momentum. The radial evolution of the star particle then proceeds inwards while the particle continues to gain angular momentum until epicycle motion resumes at $t\sim 2.06$ Gyr.

\subsection{Orbits of negative migrators}

Fig. \ref{L9l} shows the evolution of an example of each type of negative migrator found in the particle sample. Each panel is the same as in Fig. \ref{L9g}. 

\subsubsection{Type 1l negative migrator}

The first type (Type 1l) of negative migrator (blue star and solid blue line in the left panels and blue solid line in the right panels of Fig. \ref{L9l}) is the counterpart of the Type 1g positive migrator shown in Fig. \ref{L9g}. The 2nd-right panel of Fig. \ref{L9l} shows that the star particle begins to feel a positive tangential force at a time of around $t\sim 1.91$ Gyr, when the particle is moving towards the pericentre phase of orbit (3rd-right and 1st-left panel of Fig. \ref{L9l}). As the spiral arm grows in density, e.g. $t=1.944$ and $1.976$ Gyr (2nd- and 3rd-left panels of Fig. \ref{L9l}), the star particle continues to feel a strong positive tangential force accompanied by a steep negative slope in the angular momentum evolution (1st-right panel of Fig. \ref{L9l}). At $t\sim 1.98$ Gyr, the radial gravitational force and centrifugal force are temporarily balanced and the star particle stays at $R\sim 7.2$ kpc for $\sim 30-40$ Myr - a similar radial pause to that of the Type 1g migrator described in the previous section. At $t\sim 2.02$ Gyr, the radial gravitational force plus the loss in angular momentum overcome the centrifugal force, hence the star particle moves radially inward again. As the spiral arm begins to fade at around $t\sim 2.06$ Gyr, the tangential force diminishes, and the star particle resumes normal epicycle motion. Again, the star particle has remained in front of the spiral arm and continually migrated along the spiral arm.

\subsubsection{Type 3l negative migrator}

The other example star particle shown in Fig. \ref{L9l} is the negative migrator counterpart of the Type 3g positive migrator, and therefore we designate it as Type 3l (black triangle and black solid line in the left panels and black dashed line in the right panels of Fig. \ref{L9l}). This star particle originates farther from the spiral arm than Type 1l as the spiral arm begins to build up at $t= 1.912$ Gyr (top-left panel of Fig. \ref{L9l}). Therefore, the star particle begins to feel the positive tangential force (deceleration) at a time of $t\sim 1.93$ Gyr, which is later than the time at which the Type 1l negative migrator begins to lose angular momentum (2nd-left and 2nd-right panels of Fig. \ref{L9l}). At this time, the star particle is close to the apocentre phase of orbit (3rd-right panel of Fig. \ref{L9l}), and consequently moves closer to the spiral arm at $t\sim 1.976$ Gyr (3rd-left panel of Fig. \ref{L9l}). Similar to the Type 3g positive migrator, the inward direction of the angular momentum change and epicycle motion means that the star particle moves rapidly towards the centre of the galaxy (left panels of Fig. \ref{L9l}). Again, the star particle reaches the pericentre at $t\sim 2.01$ Gyr, before the angular momentum loss has ceased. This means that the centrifugal force becomes greater than the radial gravitational potential even though the star particle continues to lose angular momentum. The tangential force diminishes at $t\sim 2.07$ Gyr, when the particle resumes epicycle motion.

\subsubsection{No Type 2l negative migrator}

We could not find a negative migrator counterpart to the Type 2g positive migrator. This may be attributed to the presence of another spiral arm e.g. the one seen on the front side of the main spiral arm at around ($R$ kpc, $\theta_{fs}$) $= (8, 130)$ at $1.976$ Gyr (3rd-left panel of Fig. \ref{L9l}) relative to the main spiral arm on which we focus ($R, \theta_{fs}$)$ = (8,185)$. This is closer than the spiral arm behind the main spiral arm at ($R, \theta_{fs}$) $= (8, 265)$. The closer proximity to the main spiral of the density enhancement on the front-side of the spiral arm in comparison to the density enhancement on the back-side of the spiral may cause the motions of star particles in front of the spiral to be more influenced by neighbouring density enhancements than those behind the main spiral arm.

To mirror the Type 2g positive migrator, the Type 2l negative migrators would have begun to lose angular momentum just after they passed pericentre. In this epicycle phase, the Type 2l star particle would temporarily move away from the spiral arm (because it has a higher rotation velocity in this phase than the spiral arm), and it is possible that the over-density in front of the spiral arm ``mopped up'' these star particles, which would either reduce or change sign of the positive tangential force acting upon the star particle. This would mean that these star particles gained some angular momentum from this over-density during the time window examined. Therefore, these star particles would not be strong migrators, because they would populate a region of lower $|\Delta L_z|$ in Fig. \ref{deltaL}. In this case, they will not be selected in our strong migrator sample.

\subsection{Orbits of non-migrators}

\begin{figure*}
\begin{tabular} {l r} \hspace{-2.0mm}
 \includegraphics[scale=2.5] {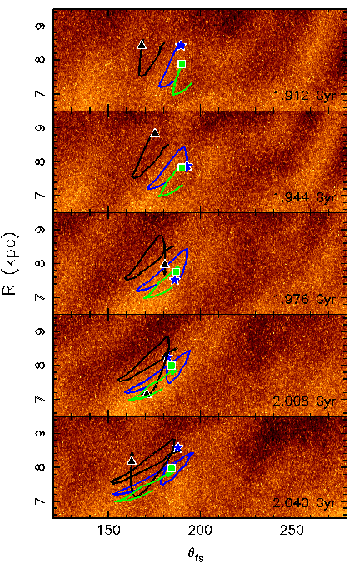} &  \includegraphics[scale=2.4] {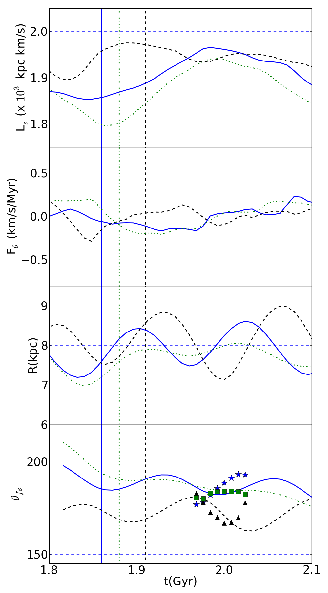} \\
\end{tabular}
\caption{The same as Fig. \ref{L9g} but for types of non-migrator.}
\label{L9n}
\end{figure*}

In this section, we analyse the orbital evolution of those star particles that experience very little or no net change in angular momentum over the time window (non-migrators). Fig. \ref{L9n} shows an example of each of the three types of non-migrator in the sample. 

\subsubsection{Type 1n non-migrator}

The first type of non-migrator (Type 1n, represented by the blue star and solid blue line in the left panels and the solid blue line in the right panels of Fig. \ref{L9n}), is very close to where the spiral arm begins to build up at $t\sim 1.912$ Gyr (1st-left panel of Fig. \ref{L9n}), and it begins to feel a negative tangential force early at $t\sim 1.86$ Gyr (2nd-right panel of Fig. \ref{L9n}). At $t=1.944$ Gyr (2nd-left panel Fig. \ref{L9n}), the star particle is located in between the main spiral arm at ($R, \theta_{fs}$) $= (8,185)$ and the weaker spiral arm behind it at ($R, \theta_{fs}$) $= (8, 205)$ - in contrast to the positive migrators which were behind both density enhancements at this time. The star particle moves towards pericentre (3rd-right panel of Fig. \ref{L9n}), while it gains angular momentum (top-right panel of Fig. \ref{L9n}). Because the star particle is around pericentre and located close behind the arm (3rd-left panel of Fig. \ref{L9n}), the star particle moves towards the spiral arm until $t\sim 1.99$ Gyr when it passes the spiral arm, as seen in the bottom-right panel and the 3rd- and 4th-left panels of Fig. \ref{L9n}. The particle is now located on the front side of the spiral and as a result the tangential force acting upon the star particle has become positive, which causes the star particle to begin to lose angular momentum. The star particle then resumes epicycle motion as the spiral arm fades at $t\sim 2.05$ Gyr. 

The general properties of these non-migrators is that they originate very close to the spiral arm, and orbit the spiral arm until it disappears, continually gaining and losing angular momentum such that the net angular momentum gain is $\Delta L_z \sim 0$, at the end of the time window. It is worth noting that when the star particle is circling around the spiral arm, the amplitude of the epicycle motion is smaller. This is because the tangential force from the spiral arm acts always to change the guiding centre in the direction opposite to that of epicycle motion. Therefore, as the star particle moves towards pericentre, it is accelerated ($F_{\theta}<0$), which increases the guiding centre and in turn increases the pericentre radius. For example, the pericentre radius is larger at $t\sim1.96$ Gyr than at $t=1.84$ Gyr, and at $t\sim2.09$ Gyr the pericentre is returned to that at $t\sim1.84$ Gyr. This indicates the amplitude of the epicycle motion is shortened, because the tangential force from the spiral arm acts against the epicycle motion. The Type 1n orbit is found to originate also in front of the spiral arm when it first feels the tangential force.

\subsubsection{Type 2n non-migrator}

The second type (Type 2n) of non-migrator (black triangle and solid black line in left panels and dashed black line in right panels of Fig. \ref{L9n}) is located in front of the spiral arm as it forms at $t \sim 1.944$ Gyr (1st- and 2nd-left panels of Fig. \ref{L9n}). It begins to feel a positive tangential force and migrates inward at $t \sim 1.91$ Gyr. This occurs just before the star particle reaches the apocentre of the orbit, and at $t\sim 1.95-1.97$ Gyr the loss of angular momentum and the inward epicycle motion (1st- and 3rd-right panels of Fig. \ref{L9n}) cause the radius of the star particle to decrease rapidly in a manner similar to Type 3l negative migrator. This causes a decrease in pericentre radius (see the radius at $t\sim1.87$ and $t\sim2.0$ Gyr in 3rd-right panel of Fig. \ref{L9n}). However, unlike the Type 3l negative migrator, this non-migrator is too close to the spiral arm when it reaches apocentre at $t=1.94$ Gyr and as a result is passed by the spiral arm at $t\sim 1.98$ Gyr (bottom-right panel of Fig. \ref{L9n}). This causes the change of tangential force from positive to negative, and the star particle then gains back the angular momentum lost previously (top-right panel of Fig. \ref{L9n}). This once again ensures that this non-migrator has a net angular momentum change of $\Delta L_z \sim 0$ at the end of the time window. 

The difference in radial evolution between this Type 2n and the Type 1n non-migrators is that the Type 2n non-migrator shows an increase in epicycle amplitude because the pericentre at $t\sim 2.0$ Gyr is lower with respect to the earlier time, $t\sim 1.86$ Gyr, i.e. the amplitude of epicycle motion is increased instead of decreased as in the case of the Type 1n non-migrator. 

It is interesting to note that this type of non-migrator is not found to originate from behind the spiral arm. This may be related to the differences between the front- and back-side of the spiral arm as discussed in Section. 5.2.3. In addition, as discussed in Section 5.1.1 and 5.3.1, there is a small density enhancement that merges with the back-side of the spiral arm at $t\sim 1.976$ Gyr. Therefore the back-side of the spiral arm seems to build up later than the front-side, and this likely causes the difference in the variety in the orbits of star particles that originate from the front- and back-side of the spiral arm.

\subsubsection{Type 3n non-migrator}

\begin{table*}
\caption{Summary of orbital characteristics for each orbital type. Columns show 1) orbital type name 2) the time at which the particle first feels the tangential force, $t_{capture}$ (measured by eye to provide an indication) 3) initial particle position with respect to the spiral arm at $t_{capture}$ 4) the epicycle phase of the particle at $t_{capture}$ 5) the effect on their epicycle amplitude 6) whether or not they remain on the same side of the spiral throughout the spiral lifetime.}
\label{tab1}
\begin{tabular}[c]{|p{1.0cm}|p{2.0cm}|p{3.5cm}|p{3.0cm}|p{2.5cm}|p{3cm}|}
\hline
Name & $t_{capture}$ (Gyr) & Initial position & Initial phase & epicycle amplitude & Remain on same side of arm \\
\hline
1g & 1.90 & behind arm & before apocentre & none & yes  \\
2g & 1.89 & behind arm & apocentre & shortened & yes  \\
3g & 1.93 & behind arm & pericentre & lengthened & yes  \\
1l & 1.91 & in front of arm &  before pericentre & none & yes  \\
3l & 1.93 & in front of arm & apocentre & lengthened & yes  \\
1n & 1.86 & close behind (in front of) arm & pericentre (apocentre) & shortened & no  \\
2n & 1.91 & in front of arm & apocentre & lengthened & no  \\
3n & 1.88 & very close behind (in front of) arm & apocentre (pericentre) & shortened & no \\
\hline
\end{tabular}
\end{table*}

The third type (Type 3n) of non-migrator particle found in this sample (the green square and solid green line in left panels and green dot-dashed line in the right panels of Fig. \ref{L9n}) begins to feel a negative tangential force at $t\sim 1.88$ Gyr (2nd-right panel of Fig. \ref{L9n}) just as it reaches the apocentre of the orbit at $t\sim 1.912$ Gyr (top-left and third-right panels of Fig. \ref{L9n}). This is accompanied by an increase in angular momentum (top-right panel of Fig. \ref{L9n}) from about $t\sim 1.88-1.99$ Gyr. During this time, the radius of the star particle proceeds to apocentre and exhibits a small dip in radius at $t\sim 1.96$ Gyr (3rd-right panel of Fig. \ref{L9n}). This is obviously the feature of an increased pericentre radius owing to the imbalance between the radial gravitational force and centrifugal force boosted by a gain in angular momentum. The pericentre-like feature is also seen in the bottom-right panel of Fig. \ref{L9n}, which shows a decreasing $\theta_{fs}$ of the star particle (line) that brings the star particle closer to the spiral arm in azimuth angle (square symbols) at $t\sim 1.96$ Gyr.

At $t\sim 1.99$ Gyr, the star particle passes the spiral arm and begins to lose angular momentum (right panels in Fig. \ref{L9n}). Up to $t\sim 1.99$ Gyr, this non-migrator exhibited similar evolution to Type 1g positive migrators, but it differed after this time by crossing the spiral arm instead of remaining behind it. This is entirely because this non-migrator is too close to the spiral arm, such that even a $v_{\theta}$ that is slightly faster than the spiral arm rotation velocity will take the particle past the spiral arm. Again this highlights the importance of azimuth angle of the star particle with respect to the spiral arm. This type of non-migrator exhibits shortened epicycle amplitude while it remains on the same side of the spiral arm. This is different from Type 2n, which shows extended epicycle amplitude, and it is also different from Type 1n, which shows shortened epicycle amplitude as it circles the spiral arm. This Type 3n non-migrator is found both in front of and behind the spiral arm, unlike Type 2n.

\subsection{The tangential force}

\begin{figure}
\begin{center}
\includegraphics[scale=0.64] {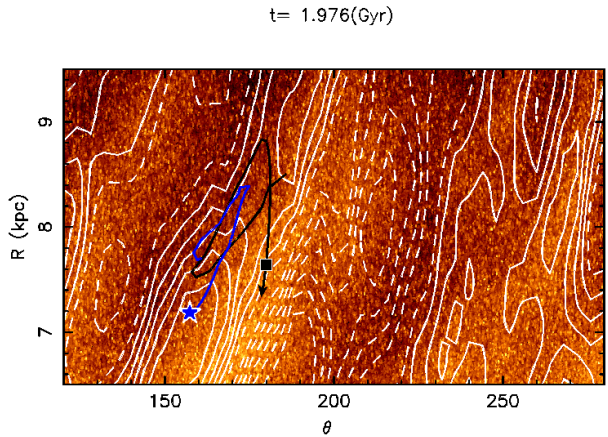} 
\caption[]{Close up of the density map in the $R-\theta_{fs}$ plane at $t=1.976$ Gyr, with contours of the tangential force over-plotted in white. Solid lines indicate positive tangential force (acting in the direction $\hat{\theta}$) and dashed lines indicate negative tangential force (acting in the direction $-\hat{\theta}$). The position of the Type 1l negative migrator and Type 2n non-migrator are represented by a blue star and black square respectively, and their orbit histories shown by the solid lines of the same colour.}
\label{tanf}
\end{center}
\end{figure}

We have shown above that time periods of sustained tangential force, which cause strong radial migration, depend on the star particle position with respect to the spiral arm. For example, the negative migrators that feel strong positive tangential forces lose angular momentum while they feel that force, which is possible because the particle always remains in front of the spiral arm. However, there exist some non-migrators that spend time in similar regions around the spiral as migrators, but upon which almost no tangential force acts.

To explore this, we compute a map of the tangential force. Fig. \ref{tanf} shows the density map of the spiral arm with contours of tangential force over-plotted in white at $t=1.976$ Gyr (solid for positive values and dashed for negative values). As expected, the tangential force is zero at the density peak of the spiral arm. The tangential force becomes stronger with increasing azimuthal distance from the spiral arm, until it reaches a maximum, $\sim10$-$20$ degree away from the density peak on either side. At farther distances from the peak of the spiral arm, the tangential force decreases and eventually changes the sign, owing to the force contributions from neighbouring spiral arms. It is interesting to see the difference between the front- and back-sides of the spiral arm. On the front-side of the arm the point of the maximum tangential force is closer to the density peak of the spiral arm than that on the back-side of the spiral arm. This is because the neighbouring spiral arm on the front-side is closer than the one on the back-side. On the back-side, the small density enhancement merges to the main spiral arm around $R=8.5$ kpc, which pushes the point of the maximum tangential force farther from the density peak of the main spiral arm.

We show the orbit history of the Type 1l negative migrator (blue star and solid blue line in Fig. \ref{tanf}) shown in Section. 5.2.1 and the Type 2n non-migrator (black square and solid black line in Fig. \ref{tanf}) shown in Section 5.3.2. The figure shows that the negative migrator is captured in a region of strong tangential force in front of the spiral arm, which continues over the spiral arm lifetime (see 2nd-right panel of Fig. \ref{L9l}). However, the non-migrator, despite being located in front of the spiral arm and having a very similar orbit to the negative migrator, feels almost no tangential force over this time period (see 2nd-right panel of Fig. \ref{L9n}) because the star particle is too close to the peak of the spiral arm. This implies that it is not enough that the star particle is located on one side of the spiral arm in order to become a migrator: the migrator must be far enough away from the spiral arm peak line - yet not too far! - in order to be captured by regions of strong tangential force such as those shown in Fig. \ref{tanf}. We therefore refine the condition for a strong positive (negative) migrator to be a star particle able to remain at a \emph{suitable} distance behind (in front of) the spiral arm. This re-emphasizes the importance of the $R\Delta\theta$ parameter in distinguishing strong migrators from non-migrators.

\section{Broader Implications of Radial migration}

The main results of this paper focus on the complicated interactions between the spiral arm and a variety of orbital types, on timescales roughly equal to the spiral arm lifetime. Each type of migrator particle experiences a change in angular momentum on these timescales (Fig. \ref{deltaL}). On longer timescales, i.e. 1 Gyr, a star particle may interact with a spiral several times. The cumulative effect of consecutive spiral arm-star particles interactions is reflected in the evolution of global properties of the disc, such as the mass and metal distribution. In this section, we briefly discuss the effects of radial migration in this simulation on longer timescales.

\begin{figure}
\hspace{-13.0mm}
\includegraphics[scale=1.65] {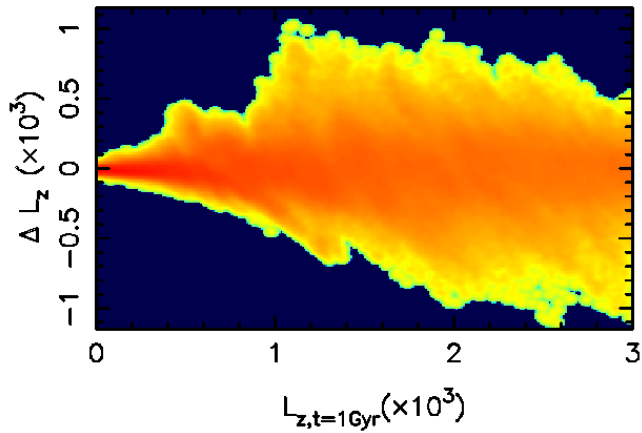} 
\caption[]{The change in angular momentum, $\Delta L_z$, between $t=1.0$ and $2.0$ Gyr of the disc particles as a function of their angular momentum at $t=1.0$ Gyr. Red colours indicate regions of high number density.}
\label{fig11}
\end{figure}

\begin{figure}
\begin{center}
\includegraphics[scale=0.45] {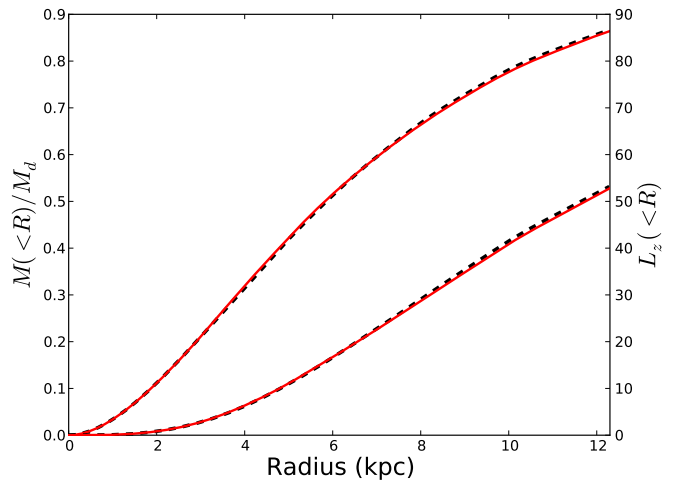} 
\caption[]{The cumulative profiles of total disc mass fraction (upper curve) and angular momentum (lower curve). The solid red and dashed black curves show the profiles at $t=1.0$ and $t=2.0$ Gyr, respectively.}
\label{fig12}
\end{center}
\end{figure}

Co-rotating spiral arms have been shown to cause radial migration at many radii, and this work confirms that migration will continue as long as the star particle remains on the same side of the spiral arm in a region of strong tangential force. This means that radial migration in this simulation is efficient at transporting star particles to different guiding radii. This is shown in Fig. \ref{fig11}, which shows the change in angular momentum of disc particles between the times of $1.0$ and $2.0$ Gyr as a function of their angular momenum at $t=1.0$ Gyr. This shows much more radial migration than for the shorter timescale shown in Fig. \ref{deltaL} ($\sim 5$ times as much). Although this radial migration redistributes the individual angular momenta of star particles (see also Figs. \ref{L9g}, \ref{L9l} and \ref{L9n}), the overall angular momentum distribution of the disc and the cumulative mass profile remain almost unchanged (see Fig. \ref{fig12}).

\begin{figure*}
\begin{center}
\subfloat{\includegraphics[scale=1.5] {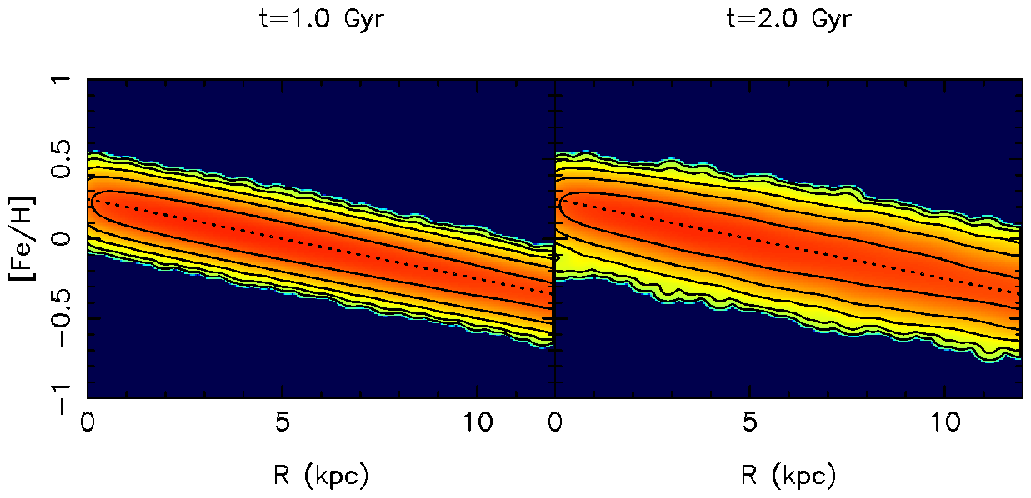}} \\
\subfloat{\includegraphics[scale=1.5] {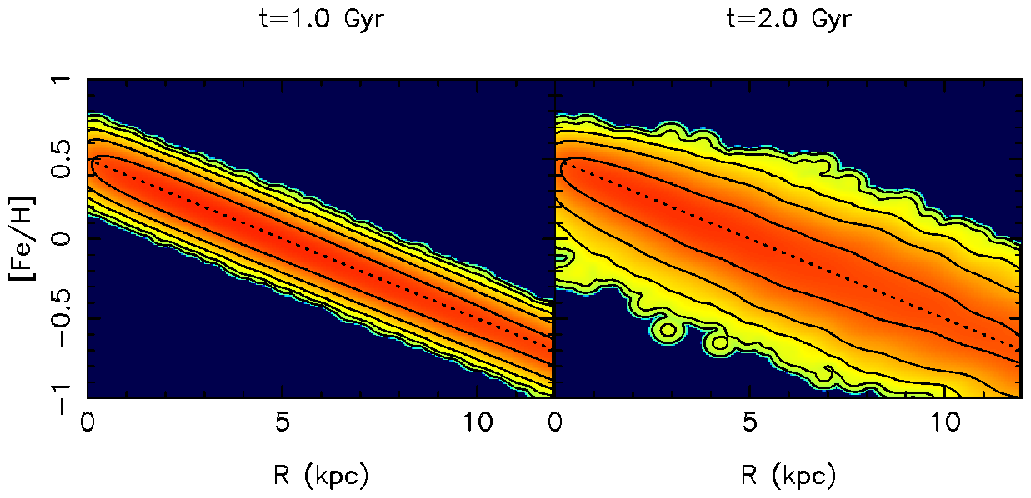}} 
\caption[]{Metallicity distribution of all disc stars as a function of radius at $t=1.0$ Gyr (left panel) and $t=2.0$ Gyr (right panel). \emph{Top:} The initial assigned metallicity gradient is $-0.05$ dex/kpc, and the dispersion of the gaussian metallicity distribution function is $0.05$ dex. \emph{Bottom:} The initial assigned metallicity gradient is $-0.1$ dex/kpc, and the dispersion of the gaussian metallicity distribution function is $0.05$ dex. In all panels the initial metallicity gradient is shown by the dashed black line. Contours and red colours indicate regions of high number density.}
\label{fig13}
\end{center}
\end{figure*}

The movement of star particles from one guiding radius to another affects the distribution of metals in the disc. To indicate the effect on the metal distribution of the radial migration in this simulation, we artificially assign metallicity values to each star particle in the simulation at a time of $t=1.0$ Gyr. We randomly assign each star particle a metal value by drawing from the gaussian metallicity distribution function at each radius with a dispersion of $0.05$ dex. The mean on which a gaussian is centred at a given radius is defined by a metallicity gradient of $-0.05$ dex/kpc, and [Fe/H]($R=0$) = 0.25 dex. The top row of Fig. \ref{fig13} shows the smoothed distribution of stars in metallicity as a function of radius at $t=1.0$ Gyr (left panel) and $t=2.0$ Gyr (right panel). The metallicity gradient does not change much at all between 1.0 and 2.0 Gyr. This is because the radial migration is not strong enough to mix stars at such a rate as to affect the slope. However, the metallicity distribution at any given radius broadens with time. The bottom row of Fig. \ref{fig13} shows the same as the top row but for an initial radial metallicity gradient of $-0.1$ dex/kpc. At $t=2.0$ Gyr, although the radial metallicity gradient remains unchanged, the metallicity distribution shows a much larger broadening owing to the the steeper gradient, as expected. Despite our crude metallicity analysis, a similar effect is seen in Fig. 16 of \citet{CSA11}, who show that in the solar neighbourhood, the metallicity distribution of stars aged between $1-5$ Gyr is comparatively broad in comparison with that of stars younger than $1$ Gyr, and both populations have the same peak metallicity value \citep[see also][]{H08}. We stress that these results are from this simulation only, and should not be taken as the general case for simulated spiral galaxies. For example, parameters such as spiral arm strength and pitch angle likely play a role in the amount of radial migration and therefore the degree of mixing that occurs in a spiral disc. Moreover, this result is derived from a N-body simulation in which metallicities have been assigned artificially at an arbitrary time. For a robust analysis on the effect of spiral morphology on metal distributions, many simulations that include the gas component and recipes for star formation and chemical evolution need to be analysed and compared with each other and observation. This topic deserves a thorough numerical study, and we reserve further discussion until then.

All of the analysis presented in this paper has been with regard to one simulation, which has a flocculent spiral structure with no bulge or bar at the centre. However, we have seen some evidence that the group characteristics of the positive, negative and non- migrators are present also in simulated galaxies of different spiral structure, and show similar separations in the phase space planes shown in Fig. \ref{phase67}. These simulations include a 3-4 armed galaxy with a static bulge in the centre \citep[simulation F in][]{GKC13}, and the barred-spiral galaxy of two arms presented in \citet{GKC12}.

\section{Conclusions}

We have performed high resolution N-body simulations of a spiral disc embedded in a static dark matter halo potential. We focus on a sample of star particles taken from around the spiral arm and divide this sample into three groups according to the amount of change in angular momentum, i.e. radial migration, of the star particles over a given time period when the spiral arm is strong. These groups are: star particles that gain the most angular momentum (positive migrators), star particles that lose the most angular momentum (negative migrators) and star particles that show almost no change in angular momentum (non-migrators).  

We follow the evolution of these groups in $v_{\theta}-v_R$, $v_{\theta}-R\Delta\theta$ and $v_{\theta}-R$ planes, and come to the following conclusions:

\begin{itemize}
\item Positive migrators tend to cluster around the region of velocity space that is slower than the mean rotation velocity (apocentre), whereas negative migrators tend to cluster around the region of velocity space that is faster than the mean rotation velocity (pericentre). Non-migrators are more evenly distributed in $v_{\theta}-v_R$ space. 
\item Positive migrators always stay behind the spiral arm ($R\Delta\theta < 0$) and negative migrators always stay in front of the spiral arm ($R\Delta\theta > 0$). Non-migrators are found on both sides of the spiral arm; those with azimuthal velocities faster than the circular velocity tend to be located behind the spiral arm, whereas those with azimuthal velocities slower than the circular velocity tend to be located in front of the spiral arm.   
\item The slope in the $v_{\theta}-R\Delta\theta$ space of the distribution of non-migrators indicates that non-migrators located behind the arm have higher azimuthal velocities the farther behind the arm they are. Similarly, the non-migrators in front of the spiral arm have lower azimuthal velocities the farther in front of the spiral arm they are when the spiral is forming.
\end{itemize}

We then examined the orbital evolution of individual star particles in each particle group in detail. We categorised and contrasted several orbital types of each group, each of which is a new type of orbit shown for the first time in this paper.

There are three types of positive migrators, each characterised by the time they feel the tangential force, their phase of epicycle motion when this force is introduced and their azimuthal distance from the spiral arm:
 
\begin{itemize}
\item Type 1g originates relatively close to the spiral arm and begins to migrate just before apocentre. It exhibits a radial pause owing to the temporary balance between radial gravitational force and centrifugal force, before continuing to move outward. 
\item Type 2g originates from a similar position to Type 1g, and begins to migrate at a similar time but after it reaches apocentre. This causes the star particle to exhibit a shortened epicycle amplitude because angular momentum increases the radius of the pericentre. 
\item Type 3g originates farther from the arm than Types 1g and 2g, and begins to migrate at later times than Types 1g and 2g, around the pericentre phase, and as a result exhibits rapid changes in radius. 
\end{itemize}

This variety of orbits shows that positive migrators can be in any epicycle phase at the time the tangential force is introduced. However, migrators that originate far from the spiral arm must be in the pericentre phase of orbit in order to catch the spiral arm and undergo large angular momentum changes. Migrators that are located closer to the spiral arm begin to migrate around apocentre because if they are in pericentre, they will pass the spiral arm and any increase in angular momentum will be arrested, as we show in Section. 5.3. Hence, the key for strong positive migrators is to stay behind the spiral arm, which is verified in the bottom-right panel of Fig. \ref{L9g} for each positive migrator example. In this way, irrespective of the phase of epicycle motion, the star particles continue to gain angular momentum from the spiral arm. We find similar types of orbits for the negative migrators. However, we did not find any Type 2 negative migrator analogue to Type 2g positive migrator. We think that this is because the spiral arm on which we focus has different properties and formation processes on the front- and back-side of the spiral arm.

We found three types of non-migrators:

\begin{itemize}
\item Type 1n originates very close to where the spiral arm forms, and always moves in the radial direction opposite to that of the change of guiding centre. They generally move toward pericentre when they are behind the spiral arm, and toward apocentre when they are in front of the spiral arm. This causes them to orbit the spiral arm for $\sim 1$ epicycle period with a shortened epicycle amplitude. This type is found on both sides of the spiral arm.
\item Type 2n originates farther from the spiral arm than Type 1n. It crosses the spiral arm from the front side as it moves towards pericentre, and therefore exhibits an elongated epicycle amplitude. This type is found on the front-side of the spiral arm only.
\item Type 3n originates close behind the spiral arm, and gains angular momentum such that it exhibits shortened epicycle motion while remaining on the same side of (behind) the spiral arm. It is so close to the spiral arm that it passes the spiral arm as it moves through pericentre. This type is found on both sides of the spiral arm.
\end{itemize}

Each type of non-migrator shows that position relative to the spiral arm is very important, and has consequences for the star particle evolution. It is again important to note that although these particles were selected as showing $\Delta L_z \sim 0$ within the selected migration time window, they may go on to migrate around other spiral arms at later times, and may have undergone migration around past spiral arms at earlier times. 

The importance of the proximity of the star particle to the spiral arm and the epicycle phase of motion of the particle found in the orbital anaylsis is consistent with the trend seen in the sample distribution in the $R\Delta\theta - v_{\theta}$ plane shown in the middle row of Fig. \ref{phase67}. It shows that strong migrators are always located on the same side of the spiral arm throughout the lifetime of the spiral arm, while non-migrators tend to pass/be passed by the spiral arm owing to the faster/slower rotation velocities of the non-migrators relative to that of the spiral arm, given close enough proximity. This underscores the criterion for star particles to become strong positive (negative) migrators as the ability of a star particle to remain behind (in front of) and stay close to the spiral arm as long as it is present.

We have found that there are certain distances from the spiral arm at which exist regions of strong tangential force. Migrators tend to be captured by these regions, whereas non-migrators may miss the strong force regions by moving too close to the spiral arm. In that case, the star particle does not feel much tangential force and does not change angular momentum. This emphasizes the importance of azimuthal distance from the spiral arm. Indeed we find that strong migrators remain at a suitable distance from the spiral arm.

The detailed information of particle orbit histories of large star samples may provide clues to how the spiral arm is formed and destroyed. This information is likely to be found in not just strong migrators and non-migrators, but from much larger star particle samples that exhibit many different degrees of radial migration. From these kind of data, it may be possible to see how the spiral arm is constructed as the influx of star particles from regions around the disc into the forming spiral arm increases the density and eventually leads to the fully formed spiral features. In the same spirit, the future trajectories of many star particles that make up the spiral features may give us an insight into how these features are disrupted. The non-linear nature of the particle motion that we have glimpsed makes this method of analysis a promising direction to pursue, and we leave this complicated investigation to future studies.

The applicability of these results to transient spirals is limited to the co-rotating spiral, given that the results hold for many radii. Therefore, this precludes the classic density wave theory, which is based on a single co-rotation radius. Furthermore, the rapid growth of spiral arms from small seeds and the persistence of spiral structure in the absence of large perturbing masses (see also D'Onghia et al. 2013) indicates the importance of highly non-linear processes in the disc that are at least not fully captured in linear approximations such as classic swing amplification theory. It may be that the spiral structure arises from swing amplification at many radii that originates at one or more radii, and amplifies and spreads to other regions of the disc when the previous amplification spills over into neighbouring radii and propagates the growth mechanism \citep[see also][]{GKC11}. We stress that this is our speculation, and note that a detailed study is required to explore other possible mechanisms of spiral arm formation. 

Finally, we remark upon the applicability of these results to real spiral galaxies. We expect that the orbital behaviour found in this paper applies to any co-rotating spiral arm, and as such, all isolated galaxies including those with a bar (Grand et al. 2012ab). However, it is possible that galaxies with companion satellites are experiencing different evolution \citep{DTP10,PB10}. Therefore, for the moment we restrict our findings to be important for isolated galaxies.

\section*{acknowledgements}
The authors thank the referee for thoughtful and constructive comments that led to the improvement of the manuscript. The calculations for this paper were performed on the UCL Legion, the Cray XT4 at Center for Computational  astrophysics (CfCA) of National Astronomical Observatory of Japan and the DiRAC Facilities (through the COSMOS consortium) jointly funded by STFC and the Large Facilities Capital Fund of BIS. We also acknowledge PRACE for awarding us access to resource Cartesius based in the Netherlands at SURFsara. This work was carried out, in part, through the \emph{Gaia} Research for European Astronomy Training (GREAT-ITN) network. The research leading to these results has received funding from the European Union Seventh Framework Programme ([FP7/2007-2013] under grant agreement number 264895.

\bibliographystyle{mn2e}
\bibliography{SPOrbits_nov19R1.bbl}

\end{document}